\documentclass[]{aa}

\usepackage{txfonts,graphicx,rotating,natbib}
\bibpunct{(}{)}{;}{a}{}{,}

\def\mps{Meteoritics~\&~Plan.~Sci.}%
\def\epsl{Earth~\&~Plan.~Sci.~Lett.}%
\def\lpscl{Lunar Planet. Sci. Conf. Lett.}%
\def\jap{J.~Appl.~Phys.}%
\def\jacs{J.~Americ.~Ceramic~Soc.}%

\begin{document}

\title{Thermal history modeling of the H chondrite parent body}

\author{Stephan Henke\inst{1} \and Hans-Peter Gail\inst{1}
 \and Mario Trieloff\inst{2} \and Winfried H. Schwarz\inst{2}
 \and Thorsten Kleine\inst{3}
}

\institute{
Universit\"at Heidelberg, Zentrum f\"ur Astronomie, Institut f\"ur Theoretische
          Astrophysik, Albert-\"Uberle-Str. 2, 69120 Heidelberg, Germany 
\and
Institut f\"ur Geowissenschaften, Universit\"at Heidelberg, Im Neuenheimer
           Feld 236, 69120 Heidelberg, Germany
\and
Institut f\"ur Planetologie, Universit\"at M\"unster, Wilhelm-Klemm-Str. 10,
            48149 M\"unster, Germany, 
  }

\offprints{\tt gail@uni-heidelberg.de}

\date{Received date ; accepted date}

\abstract
{The cooling histories of individual meteorites can be empirically reconstructed by using ages obtained from different radioisotopic chronometers having distinct closure temperatures. For a given group of meteorites derived from a single parent body such data permit the detailed reconstruction of the cooling history of that body. Particularly suited for this purpose are H chondrites because (i) all of them are thought to derive from a single parent body (possibly asteroid (6) Hebe) and (ii) for several specimens precise radiometric ages over a wide range of closure temperatures are available.
}
{A thermal evolution model for the H chondrite parent body is constructed by using the cooling histories of all H chondrites for which at least three different precise radiometric ages are available. The thermal model thus obtained is then used to constrain some important basic properties of the H chondrite parent body.
}
{Thermal evolution models are calculated using our previously developed code (Henke et al. 2012), which incorporates the effects of sintering and uses new thermal conductivity data for porous materials (Krause et al. 2011). Several key parameters determining the thermal evolution of the H chondrite parent body are varied
together with the unknown original location of the H chondrites within their parent body until an optimal fit between the radiometric age data and the properties of the model is obtained. The fit is performed in an automated way based on an `evolution algorithm'. Empirical data for the cooling history of H chondrites are taken from the literature and the thermal model is optimized for eight samples for which radiometric ages are available for at least three different closure temperatures. }
{A set of parameters for the H chondrite parent body is found that yields excellent agreement (within error bounds) between the thermal evolution model and empirical data for the cooling histories of six of the examined eight H chondrites. For two of the samples significant discrepancies exist between model and empirical data, most likely reflecting inconsistencies in the empirical cooling data. The new thermal model constrains the radius and formation time of the H chondrite parent body, and the initial burial depths of the individual H chondrites. In addition, the model provides an estimate for the average surface temperature of the body, the average initial porosity of the material the body accreted from, and the initial $^{60}$Fe content of the H chondrite parent body. 
}
{}


\maketitle


\section{%
Introduction}

Radiometric ages for chondritic meteorites and their components provide information on the accretion timescale of chondrite parent bodies, and on cooling paths within certain areas of these bodies. However, to utilize this age information for constraining the internal structure, and the accretion and cooling history of the chondrite parent bodies, the empirical cooling paths obtained by dating chondrites must be combined with theoretical models of the thermal evolution of planetesimals. Important parameters in such thermal models include the initial abundances of heat-producing short-lived radionuclides ($^{26\!}$Al and $^{60}$Fe), which are determined by the accretion timescale, and the terminal size, chemical composition and physical properties of the chondritic planetesimals. Here we use our newly developed code for modelling the structure and thermal evolution of planetesimals of $\approx100$\,km size 
\citep[][%
henceforth called paper I]{Hen11} to constrain the properties and thermal evolution of the H chondrite parent body, and to evaluate the most important parameters that determined its thermal evolution. This is done by finding the optimal fit between a theoretical thermal model and empirical data for the cooling histories of H chondrites. The H chondrite parent body was chosen here as a test case, because it is the only body for which a comprehensive set of thermochronological data is available. For eight H chondrites in particular radiometric ages exist that were obtained using at least three distinct radioisotope chronometers characterised by different closure temperatures. The cooling history of these samples, therefore, is well constrained.

\begin{table*}

\caption{Key data for cooling history of selected H chondrites}

\begin{tabular}{@{}llllllllll@{}}
\hline\hline
\noalign{\smallskip}
Meteorite & Estacado & Guare\~na & Kernouv\'e & Richard- & Allegan & Nadiabondi & Forest Vale & Ste. Mar- \\
         &          &         &            & ton      &         &            &             & guerite \\
\noalign{\smallskip}
\hline
\noalign{\smallskip}
type    & H6 & H6 & H6 & H5 & H5 & H5 & H4 & H4 & \\
\noalign{\smallskip}
         & \multicolumn{7}{c}{Hf-W$^{\rm f}$ (metal-silicate)} \\
\noalign{\smallskip}
radiometric age & $4557.3\pm 1.6 $ & & 4557.9$\pm$1.0 & 4561.6$\pm$0.8 & & & & 
4565.6$\pm$0.6& Ma\\
temperature     & 1150$\pm$75 & & 1150$\pm$75 & 1100$\pm$75 & & & & 1070$\pm50$ &  K \\
\noalign{\smallskip}
         & \multicolumn{7}{c}{Pb-Pb$^{\rm g}$ (pyroxene-olivine)} \\
radiometric age &  &  &  4537.0$\pm$1.1 &  &  & 4558.9$\pm$2.3 &  & 4564.3$\pm$0.8 & Ma\\
temperature     &  &  &  1050$\pm100$    &  &  & 1050$\pm100$    &  & 1050$\pm100$    &  K \\
\noalign{\smallskip}
         & \multicolumn{7}{c}{U-Pb-Pb$^{\rm e}$ (phosphates)} \\
\noalign{\smallskip}
radiometric age & 4559$\pm$2 & 4504.4$\pm$0.5 & 4522.5$\pm$2.0 & 4551.4$\pm$0.6 & 4550.2$\pm$0.7 & 4555.6$\pm$3.4 & 4560.9$\pm$0.7 
    & 4562.7$\pm$0.6 &  Ma\\
temperature$^{\rm a}$     & 720$\pm$50 & 720$\pm$50 & 720$\pm$50 & 720$\pm$50 
    & 720$\pm$50 & 720$\pm$50 & 720$\pm$50 & 720$\pm$50 &  K \\
\noalign{\smallskip}
         & \multicolumn{7}{c}{Ar-Ar$^{\rm e,h}$ (feldspar)} \\
\noalign{\smallskip}
radiometric age & 4465$\pm$5 & 4458$\pm$10 & 4499$\pm$6 & 4525$\pm$11 & 4541$\pm$11 & 4535$\pm$10 & 4552$\pm$8 
    & 4562$\pm$16 &  Ma\\
temperature$^{\rm b}$     & 550$\pm$20 & 550$\pm$20 & 550$\pm$20 & 550$\pm$20
    & 550$\pm$20 & 550$\pm$20 & 550$\pm$20 & 550$\pm$20 &  K \\
\noalign{\smallskip}
         & \multicolumn{7}{c}{Pu-fission tracks (pyroxene)} \\
\noalign{\smallskip}
calculated age$^{\rm c}$ & 4465$\pm$5 & 4458$\pm$10 & 4499$\pm$6 & 4525$\pm$11 & 4541$\pm$11 & 4535$\pm$10 
    & 4552$\pm$8 & 4562$\pm$16 &  Ma \\
temperature & $550\pm25$ & $550\pm25$ & $550\pm25$ & $550\pm25$ 
    & $550\pm25$ & $550\pm25$ & $550\pm25$ & $550\pm25$ & K \\
         & \multicolumn{7}{c}{Pu-fission tracks (merrilite)} \\
\noalign{\smallskip}
calculated age$^{\rm c}$ & 4401$\pm$10 & 4402$\pm$14 & 4438$\pm$10 
    & 4469$\pm$14 & 4490$\pm$14 & 4543$\pm$20 & 4544$\pm$14 
    & 4550$\pm$17 &  Ma \\
temperature & $390\pm25$ & $390\pm25$ & $390\pm25$ & $390\pm25$ 
    & $390\pm25$ & $390\pm25$ & $390\pm25$ & $390\pm25$ & K \\
\noalign{\smallskip}
         & \multicolumn{7}{c}{Pu-fission tracks (pyroxene --- merrillite)} \\
\noalign{\smallskip}
time interval$^{\rm d,e}$ & 64$\pm$9 & 56$\pm$9 & 61$\pm$8 & 56$\pm$9 & 51$\pm$9 & $>-8$  & $<8$ 
    & 12$\pm$11 &  Ma \\
cooling rate & 2.5$\pm$0.4 & 2.8$\pm$0.6 & 2.6$\pm$0.5 & 2.9$\pm$0.5 & 3.2$\pm$0.6 & (fast) & $>20$ 
    & 12.8$\pm$6 &  K/Ma \\
\noalign{\smallskip}
\hline
\end{tabular}

\medskip{\scriptsize Notes:
\\
$^{\rm a}$~\begin{minipage}[t]{.975\hsize}
see \citet{Cer91}
\end{minipage}

$^{\rm b}$~\begin{minipage}[t]{.975\hsize}
\citet{Tri03}
\end{minipage}

$^{\rm c}$~\begin{minipage}[t]{.975\hsize}
Calculated age at 390\,K from time interval between Pu-fission track (merillite, 390\,K)
and Pu-fission tracks at 550\,K (pyroxene) compared to Ar-Ar feldspar age at
550\,K
\end{minipage}

$^{\rm d}$~\begin{minipage}[t]{.975\hsize}
Time-interval for Pu-fission track cooling rate from 550--390\,K,
metallographic cooling rate 800--600\,K
\end{minipage}

$^{\rm e}$~\begin{minipage}[t]{.975\hsize}
Data from U-Pb-Pb: \citep{Goe94}, Ar-Ar: \citet{Tri03},
metallographic cooling rates: \citet{Tay87}
\end{minipage}

$^{\rm f}$~\begin{minipage}[t]{.975\hsize}
Hf-W ages are from \citet{Kle08} and were re-calculated relative to the 
$^{182}$Hf/$^{180}$Hf of the angrite D'Orbigny, which has a Pb-Pb age of 
$t =4563.4\pm0.3$\,Ma \citep{Kle12}.
\end{minipage}

$^{\rm g}$~\begin{minipage}[t]{.975\hsize}
\citet{Bou07}
\end{minipage}

$^{\rm h}$~\begin{minipage}[t]{.975\hsize}
Recalculated for miscalibration of K decay constant \citep[explanation
see text and][]{Tri03}
\end{minipage}

}

\label{TabDatChondCool}
\end{table*}

The well preserved and smooth cooling histories of many H chondrites were used to argue that the H chondrite parent body has not been subjected to catastrophic collisions during the first $\approx100$\,Ma of its history. In that case an onion shell structure develops in which the degree of thermal metamorphism is a function of depth within the parent body; material of petrologic type H6 is found in the central part, while type H3 material is located at the surface. Based on the assumption of an onion shell structure, a number of thermal evolution models were constructed that used increasingly more complex physical input \citep{Min79,Miy81,Ben96, Akr98,Tri03,Hev06,Sah07,Kle08,Har10}. All the previous thermal models of the H chondrite parent body found a radius of about 100\,km and an accretion time of about 2\,Ma after CAI formation. The success of these models in reconstruction the empirical cooling histories of the H chondrites gave much support to the hypothesis that catastrophic collisions did not significantly disturb the initial thermal structure of the H chondrite parent body. It should be noted, however, that some hints exist for the occurrence of impacts \citep[see][and references
therein]{Wit10}, but these do not appear to have resulted in catastrophic disruption. 

These previous thermal models, despite their success in reconstructing the observed empirical cooling rates for several H chondrites, fall short in providing insights into the physical processes occurring during the thermal history of a $\approx100$\,km-sized planetesimals. This motivated us to use our new thermal evolution model (described in paper I) to find the optimal fit between the empirical cooling history of the H chondrites and the thermal model. Our new thermal model applies recent data on the thermal conductivity of granular material \citep{Kra11,Kra11b} and on the compaction by cold isostatic pressing before the onset of sintering \citep{Gue09}. The modelling of the sintering process is based on the same kind of theory \citep{Rao72}
 as in \citet{Yom84}, but with some improvements. This theory does not consider more recent approaches for modelling hot pressing for technical processes \citep[e.g.][]{Arz83,Fis83,Lar96,Sto99} but appears more appropriate for the lower pressure regime relevant for asteroids. The treatment of heat conduction of the chondritic material and its sintering under pressure and at high temperature still has to be considered as rather crude, but at least is based on physical concepts that are successfully applied in other fields and in some laboratory measurements.

We do not consider a possible melting of the body, because the H chondrite parent body is thought to have remained undifferentiated. Our results will show, however, that this assumption may not be entirely valid, because the calculated central temperature of the H chondrite parent body is only slightly below the solidus temperature of chondritic material \citep[1200\,K, ][]{Fei97}. 

Our model considers more properties of the asteroidal body than traditional approaches, and so fitting of the model to the observed cooling histories is more difficult because of the much larger parameter space. Therefore we develop an automated method to perform such a fit that is based on the "evolution algorithm" that mimics the concepts that are thought to rule biological evolution in nature. This is some kind of 'intelligent' trial-and-error method by which the optimum of some quality function can be found. Its advantage is that it does not require any kind of good behaviour of the quality function. It will be shown that this method \citep[in the version described in][]{Cha95} can successfully be applied to our problem. 

For fitting the model we exclusively use meteorites for which at least three age determinations at different closure temperatures are available. These meteorites do not only constrain the slope but also the curvature of an individual cooling path at a certain location within the parent body. Within the framework of the analytic model of \cite{Miy81}, such a data set for a single meteorite would already completely determine the radius and the formation time of the body. However, in view of the presumably complex thermal evolution of meteorite parent bodies, a larger dataset for several meteorites is necessary for a reliable fit. For the H chondrites, 31 data from 8 meteorites (2 meteorites with 5 data, 3 with 4 data, 3 with 3 data) are available for this purpose. This allows determining all the important parameters of the H chondrite parent body by an optimization procedure. 

The plan of this paper is as follows: In Sect.~\ref{SectCoolHist} we describe
the available empirical data. In Sect.~\ref{SectEvoMod} we give a brief 
overview how we calculate thermal evolution models. Section~\ref{SetFitProc}
describes the optimisation method.  Section \ref{SectFitHChond} gives
our results and conclusions. The paper closes with some final remarks in
Sect.~\ref{SectConclu}.


\section{Data for cooling histories of H chondrites}
\label{SectCoolHist}

Most ordinary chondrites were affected by thermal metamorphism in their parent bodies and exhibit a range of metamorphic conditions from type 3 (unequilibrated) to type 6 (equilibrated) \citep{Dod69}. This range in metamorphic conditions probably reflects different heating and cooling histories, presumably caused by varying burial depths within the parent body \citep{Miy81,McS89,Tri03,Kle08}. The heat source for the thermal metamorphism of ordinary chondrites most probably was heating by decay of $^{26\!}$Al
and to a lesser extent also $^{60}$Fe. 

Ordinary chondrites and their components have been dated by a variety of methods \citep{Nyq09}. Al-Mg ages for chondrules from L and LL chondrites indicate that chondrule formation occurred approximately 2\,Myr after formation of Ca-Al-rich inclusions (CAIs– commonly regarded as the oldest objects formed in the solar system) \citep{Kit00,Rud07,Rud08}. The formation age of chondrules from H chondrites is less well known because no primitive H chondrite has yet been investigated with the Al-Mg system. However, a Hf-W age for the H4 chondrite Ste. Marguerite appears to date chondrule formation at $1.7\pm0.7$\,Myr after CAI formation \citep{Kle08,Kle09}, about contemporaneously to chondrule formation in L and LL chondrites. 

The ages for chondrules from ordinary chondrites mainly constrain the timescale of parent body accretion \citep{Ale05,Kle08}. In contrast, ages for equilibrated chondrites can be used to reconstruct the cooling history of ordinary chondrite parent bodies, by applying chronometers having different closure temperatures to a set of equilibrated chondrites of different metamorphic grade. By far the most comprehensive set of such age data exists for H chondrites (see Table~\ref{TabDatChondCool}), providing temporal information for cooling over almost the entire range of temperatures prevailing from the metamorphic peak down to ambient temperature. The closure temperatures for diffusive exchange of parent and daughter elements among the different minerals in a rock \citep{Dod73,Gan01} for radiometric methods applied to H chondrites range from $\approx1150$\,K (Hf-W) to $\approx550$\,K (Ar-Ar). Additional information on the cooling history is provided by $^{244}$Pu fission tracks in orthopyroxene (corresponding to cooling below $\approx550$\,K) and merrillite (corresponding to cooling below $\approx390$\,K) \citep{Pel97,Tri03}, and from metallographic cooling rates (corresponding to cooling between $\approx800$ and $\approx600$\,K), which employ Fe-Ni diffusion profiles in metal grains consisting of kamacite and taenite \citep[e.g.,][]{Her94, Hop01}. 

The age and thermochronological data available for H chondrites are summarised in Table~\ref{TabDatChondCool}, along with information on the closure temperatures of the different methods. These age data permit a detailed reconstruction of the cooling history of H chondrites, and form the basis for numerically simulating the accretion and cooling history of the H chondrite parent body (Sect.~\ref{SectFitHChond}).

For calculating heating rates the difference between the formation time of CAIs and the ages from Table~\ref{TabDatChondCool} has to be formed. This sets the zero-point of our time-scale and the instant for which the initial abundance of $^{26\!}$Al is prescribed. As reference time we take the value of
4\,568.5$\pm$0.5\,Ma from \citet{Bou07}.

\section{Thermal evolution models}
\label{SectEvoMod}

The temperature history of an undifferentiated meteoritic parent body is
assumed to be determined by essentially three processes: (i) Transient heating
by decay of short-lived radioactive nuclei $^{26\!}$Al and $^{60}$Fe and long
lasting heating by  long-lived radioactives ($^{40}$K, $^{232}$Th, $^{235}$U, 
$^{238}$U), (ii) transport of heat to the surface by heat conduction, and (iii)
energy exchange with the environment by emission and absorption of radiation
energy. The heat transport in the body is governed by the fact that the
material is porous, that heat conductivity of chondritic material strongly
depends on porosity \citep[e.g.][]{Kra11b}, and that the porosity changes by
sintering as the body is heated up. 

\begin{table}

\caption{Fixed parameters of the H chondrite parent body models (taken from
paper I).}

\begin{tabular}{l@{\hspace{.7cm}}ll}
\hline\hline
\noalign{\smallskip}
quantity & value & unit \\
\noalign{\smallskip}
\hline 
\noalign{\smallskip}
bulk density  $\rho_{\rm b}$ & 3.78 & g\,cm$^{-3}$ \\
\noalign{\smallskip}
mass fraction $X_{\rm Al}$ & $9.10\times10^{-3}$ & \\
mass fraction $X_{\rm Fe}$ & $2.93\times10^{-1}$ & \\
$^{26\!}$Al/\,$^{27\!}$Al ratio (at CAI formation) & $5.1\times10^{-5}$ & \\
\noalign{\smallskip}
\hline
\end{tabular}

\label{TabFixedParm}
\end{table}

For the formation of the bodies we apply the ``instantaneous formation'' 
approximation, i.e., it is assumed that the bodies acquired most of their mass
during a period short compared to the half-life $\tau_{1/2}=7.2\times10^5$\,a of
the main heating source, $^{26\!}$Al. Present planetary formation scenarios
suggest such rapid mass acquisition of 100\,km sized and bigger bodies 
\citep[e.g.][]{Wei06,Naga07}. The rapid growth is approximated by the 
assumption that the bodies came into life at some instant $t_{\rm form}$ and
have constant mass over their subsequent evolution. 

For reasons of computational economics the shape and internal structure of the
body is assumed to be spherically symmetric, though this is certainly a strong
simplification. In detail, the following equations are solved: 

1. The equation of hydrostatic equilibrium
\begin{equation}
{{\rm d}\,p\over{\rm d}\,r}=-{GM_r\over r^2}\,\varrho\,,
\label{EqHydro}
\end{equation}
for the pressure $p$, where
\begin{equation}
M_r=4\pi\int_0^r{\rm d}r'\,r'^2\varrho (r')\,.
\label{DefMr}
\end{equation}
and 
\begin{equation}
\varrho=\varrho_{\rm b}D\,.
\end{equation}
The density $\varrho_{\rm b}$ is the mass-density of the consolidated material 
and $D$ the fraction of volume filled  with matter. The void fraction or 
porosity of the material is $\phi=1-D$.

2. The heat conduction equation for the evolution of the temperature
\begin{equation}
c_v{{\rm d}\,T\over{\rm d}\,r}+{1\over \varrho r^2}
{\partial\over\partial\,r}\,r^2q_r=+h
-{P\over\rho}{{\rm d}\over{\rm d}\,t}\,{1\over\varrho}+v_r F_r\,,
\label{EqT0}
\end{equation}
where $c_v$ ist the specific heat of material per unit mass, $h$ the specific
heating rate by decay of short living radioactives, $v_r F_r$ the local 
gravitational acceleration and $v_r$ the shrinking velocity during sintering. 
The heat current is
\begin{equation}
q_r=-K{\partial\,T\over\partial\,r}
\end{equation}
with the heat-conduction coefficient $K$.
 
The heat conductivity used in our model depends on the porosity 
$\phi=1-D$ of the material. We use the analytic fit 
\begin{equation}
K(D)=K_{\rm b}\left({\rm e}^{-4(1-D)/\phi_1}+{\rm e}^{4(a-(1-D)/\phi_2)}\;
\right)^{1/4}\,.
\label{HeatCond}
\end{equation}
to experimental data discussed in \citet{Kra11} and in paper I, where 
$\phi_1=0.08$, $\phi_2=0.167$, and $a=-1.1$ are constants given in paper I.%
\footnote{In paper I the constant $a$ was unfortunately given without the minus sign}
The pre-factor $K_{\rm b}$ corresponds to the heat conductivity of the bulk 
material (i.e., at  $D=1$). The average value of $K_{\rm b}$ is determined by
extrapolating data obtained from a couple of ordinary chondrites \citep{Yom83}
by means of Eq.~(\ref{HeatCond}) to zero porosity. Its value  is 
$K=4$\,W\,m$^{-1}$s$^{-1}$, but there is significant scatter.

3. An equation for the sintering of the initially porous material under the
influence of pressure 
\begin{equation}
{\partial\,D\over\partial\,t}=F(D,p,T)\,,
\end{equation}
where the rhs. is determined by solving a set of equations for the specific 
sintering model of \citet{Kak67} and \citet{Rao72}.
 
The internal structure and thermal evolution of parent bodies of planetesimals
is calculated by the numerical method described in paper I.

The radius of the body is not fixed but changes during the evolution by
sintering of the initially porous material. If the body would be completely
sintered the final radius would be $R$. This radius is related to the mass by
$M=(4\pi/3)\rho_{\rm b}\,R^3$ with $\rho_{\rm b}$ being the bulk density
of the completely consolidated material. Initially the material is porous and the 
radius is bigger because the density of the porous material $\rho=D\rho_{\rm b}$ is lower. The initial porosity is~$\phi_{\rm srf}$. This equals also the porosity of the outermost surface material which is too cool and under too low pressure for sintering.

The problem depends on a number of parameters, the most important of 
which are the bulk density $\rho_{\rm b}$ of the chondritic material, 
the mass $M$ of the body  or its radius $R$, the formation time $t_{\rm form}$ (measured relative to the formation time of CAIs), the heat conduction properties of the material ($K_{\rm b}$), the mass fractions $X_{\rm Al}$ and
$X_{\rm Fe}$ of Al and Fe, respectively, in the  chondritic material and the 
quantities of short-lived radioactives $^{26\!}$Al/\,$^{27\!}$Al and $^{60}$Fe/\,$^{56}$Fe the body is endowed with, the porosity $\phi_{\rm srf}$ of the 
surface material, and the surface temperature~$T_{\rm srf}$. 

The surface temperature $T_{\rm srf}$ depends during the initial phase of evolution on the properties of the ambient dusty accretion disk, after disk dispersal on the distance to the proto-sun. Since model properties depend only
weakly on $T_{\rm srf}$, this quantity is held fixed over the whole evolution period in this study.  

Some of the parameters are fixed by the properties of the H chondrite material
and general properties of the solar system. They are given in 
Table~\ref{TabFixedParm} and their values are chosen as in paper~I. The
remaining parameters $M$ or $R$, $t_{\rm form}$, $T_{\rm srf}$, 
$\phi_{\rm srf}$, and $K_{\rm b}$ are special properties of individual bodies
and are varied in our model calculations.

Since there seem to be doubts that the $^{60}$Fe/\,$^{56}$Fe ratio (at time of
CAI formation) is the same for all bodies \citep[see][]{Qui10} we prefer to
consider also this quantity as a free parameter (details are given in Sect.~\ref{SectFitParm}).

\begin{figure}[t]

\includegraphics[width=\hsize]{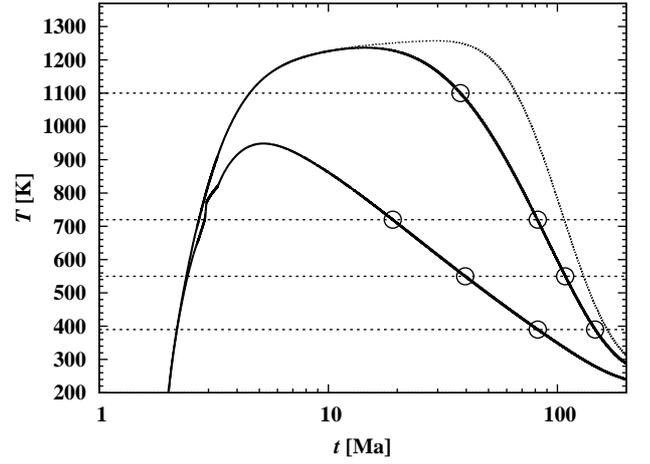}

\caption{Schematic representation of the temperature evolution at two 
different depths (full lines) and at the centre (dotted line) of a planetesimal
of the 100\,km size class. The horizontal dashed lines correspond to fixed 
temperatures equal to the closure temperatures in Table \ref{TabDatChondCool}.
The circles indicate the instants at which at a given depth the temperature
drops below a given temperature. These instants correspond to the closure 
ages of the different thermochronometers used for meteorites.
}

\label{FigTestMod}
\end{figure}

\section{%
The fitting procedure}
\label{SetFitProc}

\subsection{%
Comparing thermochronological data with asteroid models}

The data on the cooling history of meteorites described in 
Sect.~\ref{SectCoolHist} associate with a meteorite a number of 
temperature-time data pairs, corresponding to the instants where the temperature
at the pristine location of the meteorite inside of its parent body fell below
that temperature. The set of data pairs obtained for a meteorite corresponds to
a set of intersection points between the temperature evolution curve of some
mass-element with the lines of constant temperature on the descending part of the
evolution curve. This requires, of course, that the temperature evolution at the
location of the later meteorite proceeds without substantial disturbance by 
external influences (impacts, for instance).

Figure~\ref{FigTestMod} demonstrates this for the four closure temperatures
(1150\,K, 720\,K, 550\,K, 390\,K) of Table~\ref{TabDatChondCool}. The 
cooling curves are shown as full lines, 
the lines of constant temperature as dashed horizontal lines. The circles on a
cooling curve correspond to the data for a hypothetical meteorite. Since the
average slope of the curve connecting two neighbouring data points 
significantly varies for curves corresponding to different depth's, it is 
obviously possible to reconstruct for a given model from only two data points
the appropriate depth for which the temperature evolution curve runs through
just these two data points.

If the properties of the body are not known and only two data points are
available for a meteorite, such a fit would be possible for a whole range of 
model parameters. This does not allow any conclusion on the properties of the
body. If, however, more than two data points are available for a
meteorite this restricts the possible models since the overall shape of the
temperature evolution curves depends  on the model properties. The average 
slope of that part of the curve that connects neighbouring data points on the 
curve varies for different such pairs with model properties in a different
way. This allows to reconstruct the properties of the body from cooling data
on meteorites by looking for a set of model parameters for which for each of the
meteorites an appropriate depth can be found such that the cooling curve runs
through all data points. 

Since for real meteorites, experimentally determined closure temperatures 
and ages are subject to considerable errors (cf. Table~\ref{TabDatChondCool}),
data sets for an as big as possible number of meteorites (with at least three
data points) are required and the task is to find a set
of parameters that describe the properties of the parent body such that for
each of the meteorites a depth below surface can be found such that the 
(somehow defined) minimum distance of all of the data points of a meteorite
from the corresponding temperature evolution curve becomes as small as possible;
and this should hold for all available meteorites. Hence one has to solve
a minimisation problem, where some quality function has to be minimised with
respect to a set of data and a set of parameters.
 
\begin{table}

\caption{Parameter space of variable parameters for the meteoritic parent bodies}

\begin{tabular}{lll}
\hline\hline
\noalign{\smallskip}
quantity & value & unit \\
\noalign{\smallskip}
\hline
\noalign{\smallskip}
Formation time $t_{\rm f orm}$ &  1.5 -- 3 & Ma  \\
Radius $R$ &  0 -- 200 & km \\
$^{60}$Fe/\,$^{56}$Fe ratio & 0 -- $2.0\times10^{-6}$ & \\
Initial/surface temperature $T_{\rm srf}$ & 150 -- 400 &  K\\
Surface porosity $\phi_{\rm srf}$ & 0.2 -- 0.8 & \\
bulk heat conductivity $K_{\rm b}$ & 1.0 -- 5.0 & W\,m$^{-1}$\,K$^ {-1}$ \\ 
\noalign{\smallskip}
\hline
\end{tabular}

\label{TabVarParm}
\end{table}

{\ bf In the past, this task has been solved mainly by trial and error. Formation}
times $t_{\rm form}$ and parent body radii $R$ for some bodies have been 
determined this way \citep{Miy81,Ben96,Akr98,Tri03, Hev06,Sah07,Kle08,Har10}.
However, the essential properties of an asteroidal body that determine its
thermal evolution, are not merely $R$ and  $t_{\rm form}$ but also some
additional parameters, see Table~\ref{TabVarParm}. To these, one has to add the 
unknown depth's for each meteorite as additional parameters of the optimisation
problem. Here we aim to solve the problem by a general numerical algorithm that
allows to treat the more extended parameter space (and possibly additional
parameters in future calculations) and large data sets.  

\subsection{The data set}

For reconstructing the parent body of a group of meteorites that are thought
to derive from the same asteroidal body, we use all specimens for which at least
three thermochronological data points are available. For the parent body of,
e.g., the H chondrites, the eight meteorites given in Table~\ref{TabDatChondCool}
have sufficiently accurate determined data and the required minimum number of data points. Three of them have even four data points, the other five have three
of them. 

The burial depth of each of the $J$ meteorites, enumerated by an index $j$
$(=1,\dots,J)$ , in the parent body is characterised by the $J$ 
mass-coordinates $M_j$ in the theoretical model of the parent body, because the
model calculation uses the mass-coordinate as independent variable 
(see paper I). Then we have a list of $I$ data quintuples
\begin{equation}
\varpi_i=\left(T_i,\sigma_{T,i},t_i,\sigma_{t,i},j_i\right)ß,\quad(i=1,\dots,I)
\end{equation}
with $T_i$ being the closure temperature, $t_i$ the corresponding age, $\sigma_
{T,i}$ and $\sigma_{t,i}$ the errors of $T_i$ and $t_i$, respectively, and
$j_i$ a pointer to the appropriate member in the list $(M_j, j\in\{1,\dots,J\})$
of mass-coordinates corresponding to the different meteorites. 

The mass-coordinates $M_j$ are not known {\it a priori\/}. They have to be 
determined simultaneously with the parameter set given in Table~\ref{TabVarParm}
by the optimisation process.
 
\subsection{The quality function}

The quality function to be optimised can be defined in different ways. We
construct in this paper one in the following way: For each of the 
mass-coordinates $M_j$ it is known which of the data sets $\varpi_i$ are
associated with this $M_j$. During a model run, the instants when the temperature
drops below the closing temperatures belonging to the data sets $\varpi_i$
associated with $j$, are recorded. This defines for each $i$ an age $t_{{\rm
mod},i}$ on the temperature evolution curve $j$ where it passes through the
prescribed temperature $T_i$. Analogously, the temperature $T_{{\rm mod},i}$ is
recorded at instant $t_i$ for each of the data points. 

As quality function for the optimisation process we choose the least squares deviation of the model age $t_{{\rm mod},i}$ from the empirically determined age $t_i$, and of the model temperature $T_{{\rm mod},i}$ from the closure 
temperature $T_i$:
\begin{eqnarray}
\Phi(\varpi)= \sum_{i=1}^I \left(\frac{t_i-t_{{\rm mod},i}}{\sigma_{t,i}} \right)^2
+\sum_{i=1}^I \left(\frac{T_i-T_{{\rm mod},i}}{\sigma_{T,i}} \right)^2
=\chi^2\,,
\label{QualFuncMet}
\end{eqnarray}
This $\chi^2$ is to be minimised. The parameters ($R$, $t_{\rm form}$,
$^{60}$Fe/$^{56}$Fe, $T_{\rm srf}$, $\phi_{\rm srf}$, $K_{\rm b}$, $M_j$ ($j=1,\dots,J)$) are
then determined such that each cooling curve corresponding to the burial depth of
a certain meteorite in the model passes as close as possible to the empirically
determined age-closure temperature pairs for all data points of that meteorite.

\subsection{Choice of the optimisation algorithm}

The basic problem is to determine the absolute minimum of the quality function
$\Phi(\varpi)$ and possibly all other deep local potential minima with
comparable depth. As is well known, such an optimisation problem with a large
number of variables cannot be solved by analytical methods and the task of
solving the problem by numerical methods becomes extremely difficult and time
consuming if the surfaces $\Phi(\vec \varpi)=const$ have a complicated
geometric structure, which is expected to be the case for our problem. No general method is known which allows to definitely

\begin{enumerate}

\item find \emph{all} local extremal values of an optimisation problem and to

\item decide if the real optimum indeed is contained within the set of extremal values
found by the applied method.

\end{enumerate}

\noindent
Methods have been developed which allow one to find an in a certain sense 
\emph{best} result, which means that for test cases with a known absolute
minimum the algorithm always finds this minimum or at least a local minimum that
is not much worse than the absolute one.

One method, that has successfully been applied to several difficult
optimisation problems like that of the travelling salesman is the ``evolution
algorithm'' developed by \citet{Rec73, Rec89}. A detailed description of how
this method can be applied to problems in astronomy and astrophysics and a code
are given in \citet{Cha95}. We use this method to reconstruct the parent body's
properties of a suite of meteorites.

The basic idea is to imitate the concepts that are thought to rule the 
evolution of species in an algorithm to find the extremal values of some quality function.
Evolution theory describes the change of the shape of creatures in a population
during many generations due to the variability of the population and the
dependence of the amount of the creature's offspring on their conformity to the
environment. The biologic term genotype means the whole heritable attributes of
a creature, while phenotype denotes its appearance. During the development of
the creature the phenotype forms according to the instructions of the genotype.
The position in the genotype, where a certain property (e.g. eye colour) of the
individual is encoded, is called gene. The value of the property (e.g. blue) 
is called allele. The amount of possible offspring (the reproduction outcome)
depends on, how advantageous its alleles are. The reproduction outcome
determines, how numerously certain alleles in a population are inherited
compared to other respective alleles. Which processes are advantageous is
determined by the environment. Thus after some generations advantageous alleles
will accumulate in the population, while less advantageous ones will be
depleted, such that after a certain number of generations only those alleles,
the creature benefits most from, remain in the population in a noteworthy
amount. This leads to a degeneration of the gene pool. But additionally
mutations can cause random change of the genotype in a certain allele, which
can cause change of the affected gene in a significant way and can create a
completely different phenotype. Thus mutation adds novel alleles to the
gene pool. These can be superior to the established alleles (even if they are
destructive in most cases). So the mutations keep the variability of a
population, and allow the population to adapt to changing environments.   

An algorithm for finding parameters that works like evolution due to
inheritance and natural selection is superior in finding of optimal parameters
in ugly problems with many local minima compared to methods that, e.g.,
calculate in some way the derivative of the quality function and then go into 
the direction of downward slope and repeat calculating the derivative and so on.
Those methods readily find the next local minimum which, however, might not be
the absolute one. In the evolution algorithm mutations can produce extensive
changes of the parameters to the parents' ones. This can cause big jumps in the 
parameter space. Thus the population will not remain in a local minimum, but
test remote regions again and again and perhaps find a deeper local minimum that
is possibly a total one.

\subsection{General implementation of an evolution algorithm}

An evolution algorithm can be applied to every problem in which the variables
to be optimised can be encoded to form a string of numbers (chromosome). This
string then is the genotype. It is assumed in the following that for every
individual the model is calculated and compared to the data by calculating 
a quality function, like Eq.~(\ref{QualFuncMet}), e.g., and ranked by this to
other individuals. After determining which individuals are to get off-springs,
their genotypes are cut at the same random digit(s) and the related parts are
exchanged, so that, e.g., the initial part of parent 1 is connected to the end
part of parent 2. The so generated string then represents the genotype of the
new offspring individual. Mutation can be performed by giving every digit of
the string of the genotype a certain possibility to take a random value.

An implementation of an evolution algorithm can be as follows:
\begin{enumerate} 
\item A random initial population is constructed with normally no reference to
the problem. 

\item The fitness function of each individual is calculated by a user given
algorithm which determines how well the individuals match the problem.

\item A new generation is constructed by breeding of selected individuals of
the old population. There the algorithm creates new parameters by exchange of
information between the parameters of the two parent individuals. Also
the mutation takes place in this step. In most cases the number of individuals
in each generation will remain the same.

\item The old population is replaced by the new one.

\item Then back to 2. 
\end{enumerate}

\subsection{Application to our problem}
\label{SectApplOpt}

The particular implementation of a genetic algorithm that we use is that described in \citet{Cha95}. It uses a random number generator that generates
numbers between 0 and 1 to derive a set of random parameter values for
formation time, radius and all other parameters of the model. Each time
the genetic algorithm requires a new value of $\Phi$ for some individual it
provides a set of $K$ random numbers $0\le\xi_k\le 1$ ($k=1,\dots,K$) from which
values for the corresponding physical parametres $x_k$ of the model are
calculated as
\begin{equation}
x_k=x_{k,\min}+(x_{k,\max}-x_{\min})\,\xi_k\,.
\end{equation}
Here $x_{k,\min}$ and $x_{k,\max}$ are the minimum and maximum allowed values
for $x_k$ according to Table~\ref{TabVarParm}. For this set of parameter 
values $x_k$ an evolution  model is calculated and from its results the
corresponding value of the quality function $\Phi$.

The burial depth's of the meteorites are treated as part of the parameter set
of the model. Random values are assigned to the burial depths of each of the
distinct H-chondrites. In this way for  each of the individuals of the
evolution algorithm, a certain depth is ascribed  to every chondrite. 
Technically this is done by assigning to each meteorite a layer characterized
by an index $i$ and a position factor $y$ that gives its relative position
inside of the layer with respect to the edges of the layer, $r_i$ and 
$r_{i-1}$, so that the real position of the meteorite is
\begin{equation}
r_{\rm met} = r_{i-1} + y\cdot(r_i - r_{i-1})\,.
\end{equation}
Because the sintering behaviour is treated as the same everywhere within a
single layer the factor $y$ will not change during the thermal evolution. The
temperature at the position of the meteorite is calculated by 
\begin{equation}
T_{\rm met}= T_{i-1}+ y \cdot(T_i - T_{i-1})
\end{equation}
and it is tested how good the temperature evolution of that $T_{\rm met}$
matches the empirical data of the corresponding meteorite by means of
Eq.~(\ref{QualFuncMet}).

According to \citet{Cha95} a fit achieved by a genetic algorithm is moderately good if $I-K \backsim \chi^2$, where $I$ is the number of data points to be fitted and $K$ the number of the optimised parameters. For example, for the six parameters given in 
Table~\ref{TabVarParm}, plus the eight layers of
the eight meteorites ($K=6+8=14$), and the 28 data points, one gets a desired
value of $\chi^2 < 14$ for an acceptable fit.

\section{%
Results for the parent body of H-chondrites}
\label{SectFitHChond}

\subsection{Parameters}
\label{SectFitParm}

Besides the physical parameters of the problem, that are not subject of the
fitting procedure, we have a number of parameters that describe properties of
the body and external quantities. These are given in Table~\ref{TabVarParm},
including the range of variation admitted during the optimisation process. The
following parameters require some comments:

\emph{Radius:} There are a number of reasons \citep{Gaf98,Mor94} why asteroid 
(6)~Hebe is thought to be the parent body of the H chondrites and the IIE iron
meteorites. Some general characteristics of (6)~Hebe are listed in 
Table~\ref{Tab6Hebe}. If (6)~Hebe is only a fragment of the original H chondrite
parent body, then the size of (6)Hebe provides a firm lower limit for the size
of H chondrite parent body. Hebe is a somewhat irregular shaped body 
\citep{Tor03} with dimensions $205\times185\times170$\,km corresponding to an
average diameter of 186\,km. Its high mass-density means that it is not a rubble
pile as most other asteroids. It did not suffer catastrophic collisions that
disrupted the whole body and where the debris later re-assembled to the present
body. Nevertheless, a surface layer of unknown thickness has certainly been
eroded over time by collisions with other bodies and the present size of Hebe
defines only a lower limit to the size 4.56\,Ga ago. But this lower limit should
be observed in the model construction.

\begin{table}

\caption{Some properties of asteroid (6)~Hebe.}

\begin{tabular}{lllc}
\hline\hline
\noalign{\smallskip}
quantity & value & dimension & source\\
\noalign{\smallskip}
\hline
average radius $R$ & 93 & km & a \\
total mass $M$ & $1.28\,10^{22}$ & g & b \\
average mass-density $\varrho$ & 3.81 & g\,cm$^{-3}$ & b \\
\noalign{\smallskip}
\hline
\noalign{\smallskip}
radius of orbit $a$ & 2.462 & AU & c\\
excentricity $e$ & 0.23 & & c \\
inclination $i$ & 14.8\degr & & c \\
\noalign{\smallskip}
\hline
\end{tabular}

\medskip\noindent{\scriptsize
Sources: (a) \citet{Tor03}, (b) \citet{Bae11}, (c) \citet{Gaf98} 
}
\label{Tab6Hebe}
\end{table}

$^{60}$Fe/\,$^{56}$Fe-\emph{ratio:} The decay of $^{60}$Fe may have contributed
to the heating of the body, in particular since the recently re-determined
half-life of  $\tau_{1/2}=2.62\pm0.04$\,Ma \citep{Rug09} significantly exceeds
the half-life of $^{26\!}$Al of $\tau_{1/2}=0.72$\,Ma. Though there is less 
$^{60}$Fe than $^{26\!}$Al, the heating period can be significantly 
prolonged in models with $^{60}$Fe compared to models with only $^{26\!}$Al
included. 

It has been questioned \citep[cf.][]{Dau08,Qui10} that $^{60}$Fe was
homogeneously distributed throughout the solar system. If true, this would mean
that the initial $^{60}$Fe abundance in the parent body is not known \emph{a 
priori.} On the other hand, a recent study by \citet{Tan12} argued for a
homogeneous distribution of $^{60}$Fe on a relatively low level of 
$1.08\pm0.21\times10^{-8}$ for the initial $^{60}$Fe/\,$^{56}$Fe abundance
ratio at time of CAI formation. This is much lower than values previously
reported, although the authors do no comment on the credibility of earlier
studies.

At such a low abundance heating, by $^{60}$Fe is nearly negligible. Because of
the presently unclear situation with respect to which value should be taken for
the $^{60}$Fe/\,$^{56}$Fe abundance ratio we treat this as an additional free
parameter for the optimisation process. The range of $^{60}$Fe/\,$^{56}$Fe
ratios considered (see Table \ref{TabVarParm}) is between no $^{60}$Fe and the
maximum reported value  \citep{Bir88}.

\begin{figure*}[t]

\centerline{
\includegraphics[width=0.31\hsize]{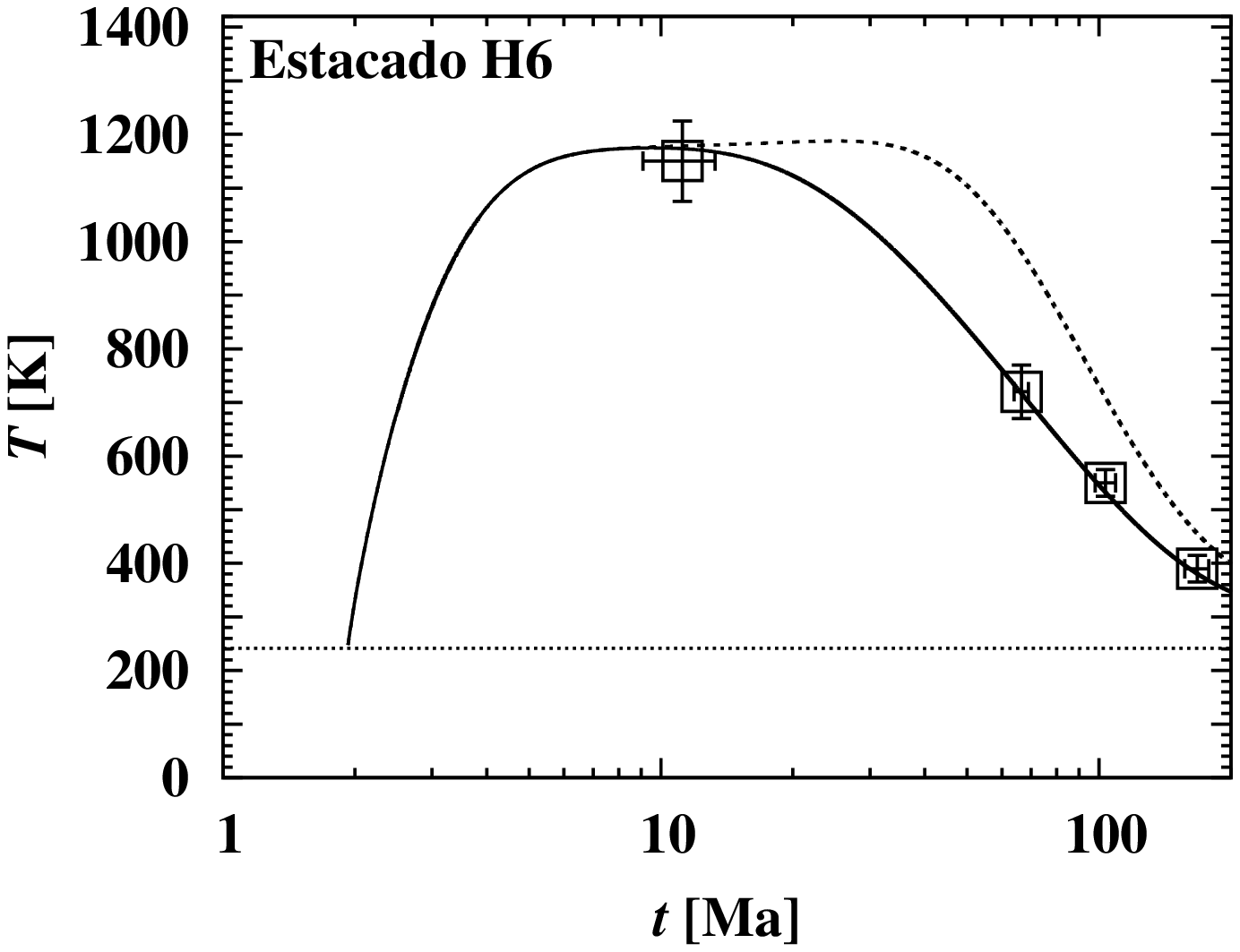}
\hfill
\includegraphics[width=0.31\hsize]{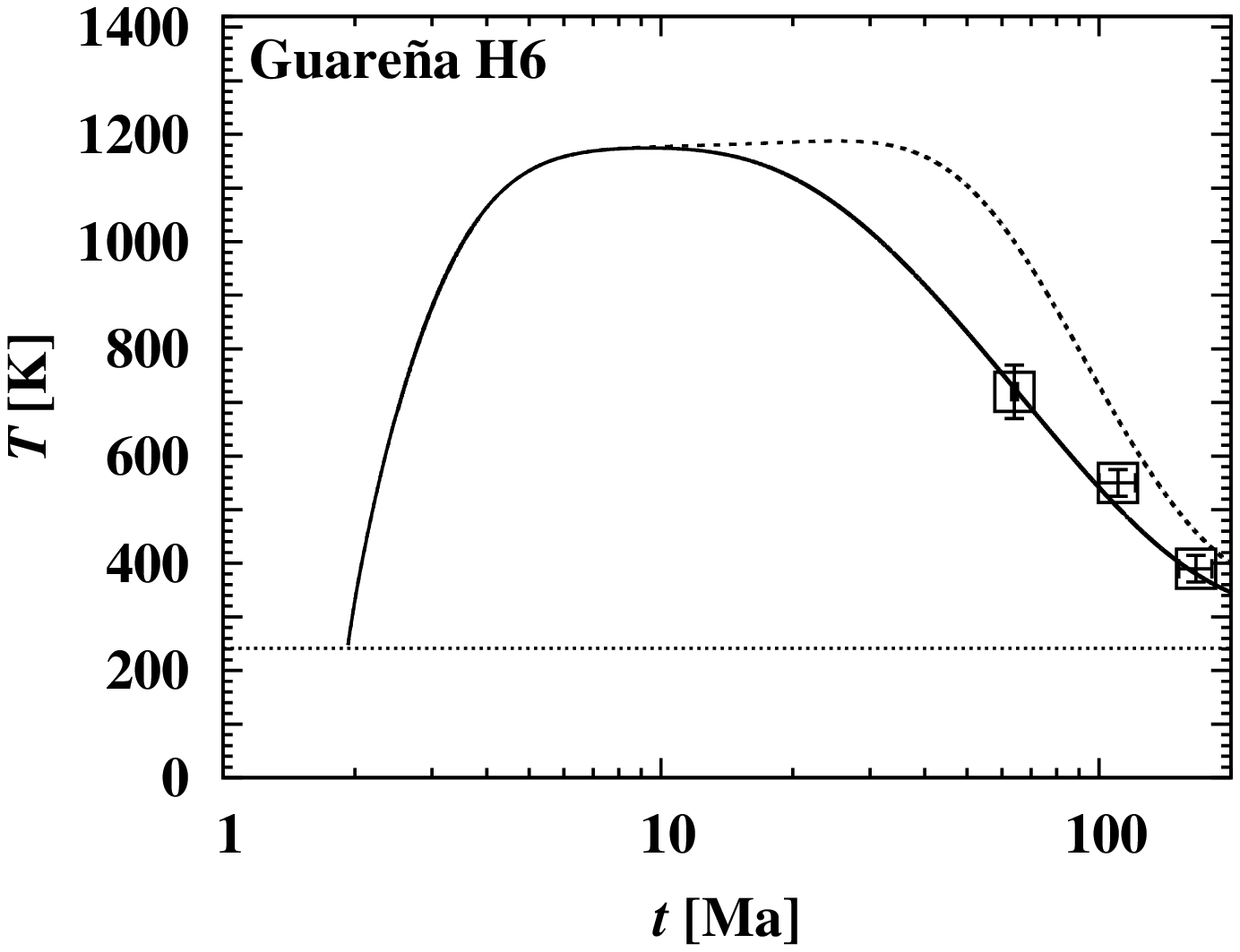}
\hfill
\includegraphics[width=0.31\hsize]{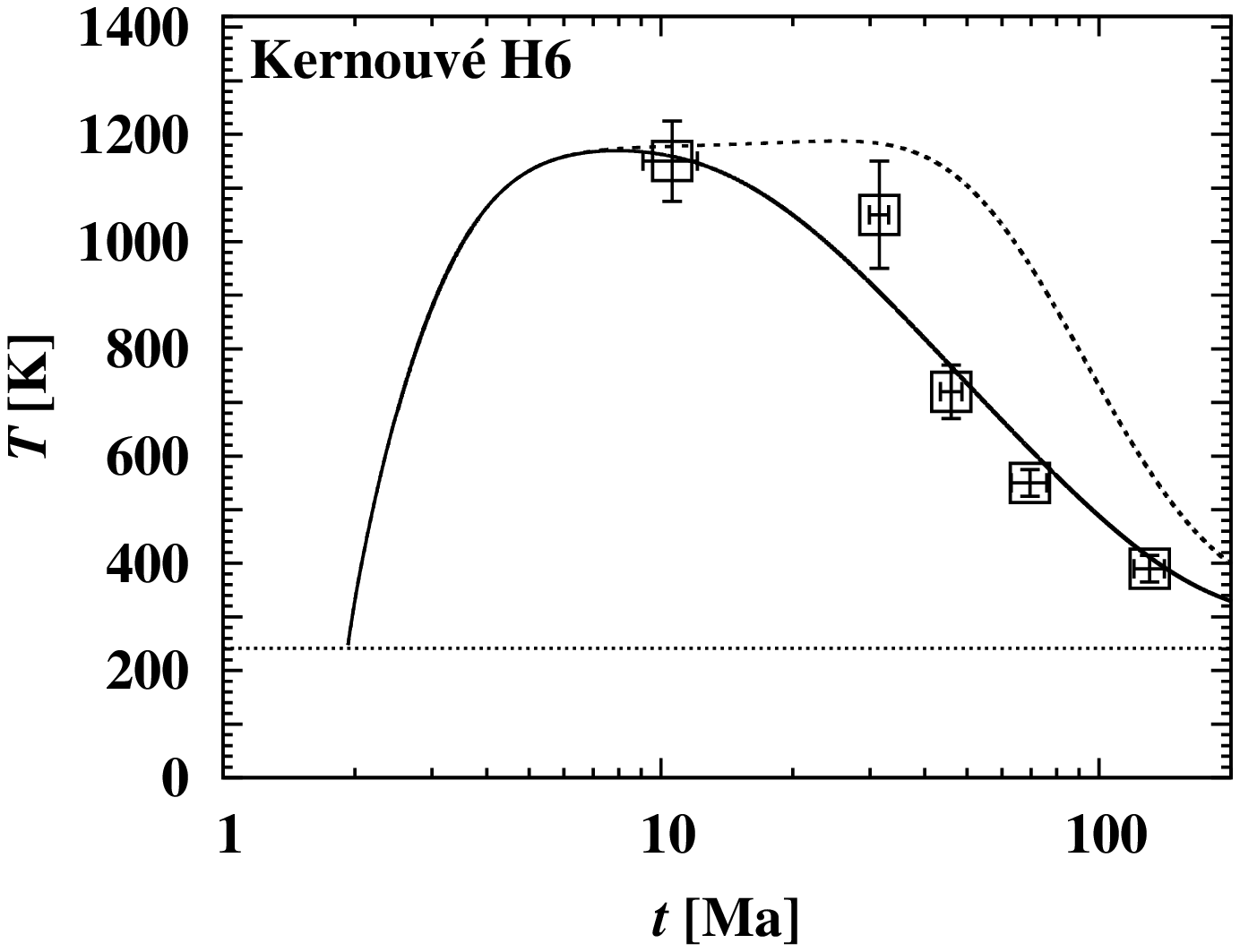} 
}

\medskip
\centerline{
\includegraphics[width=0.31\hsize]{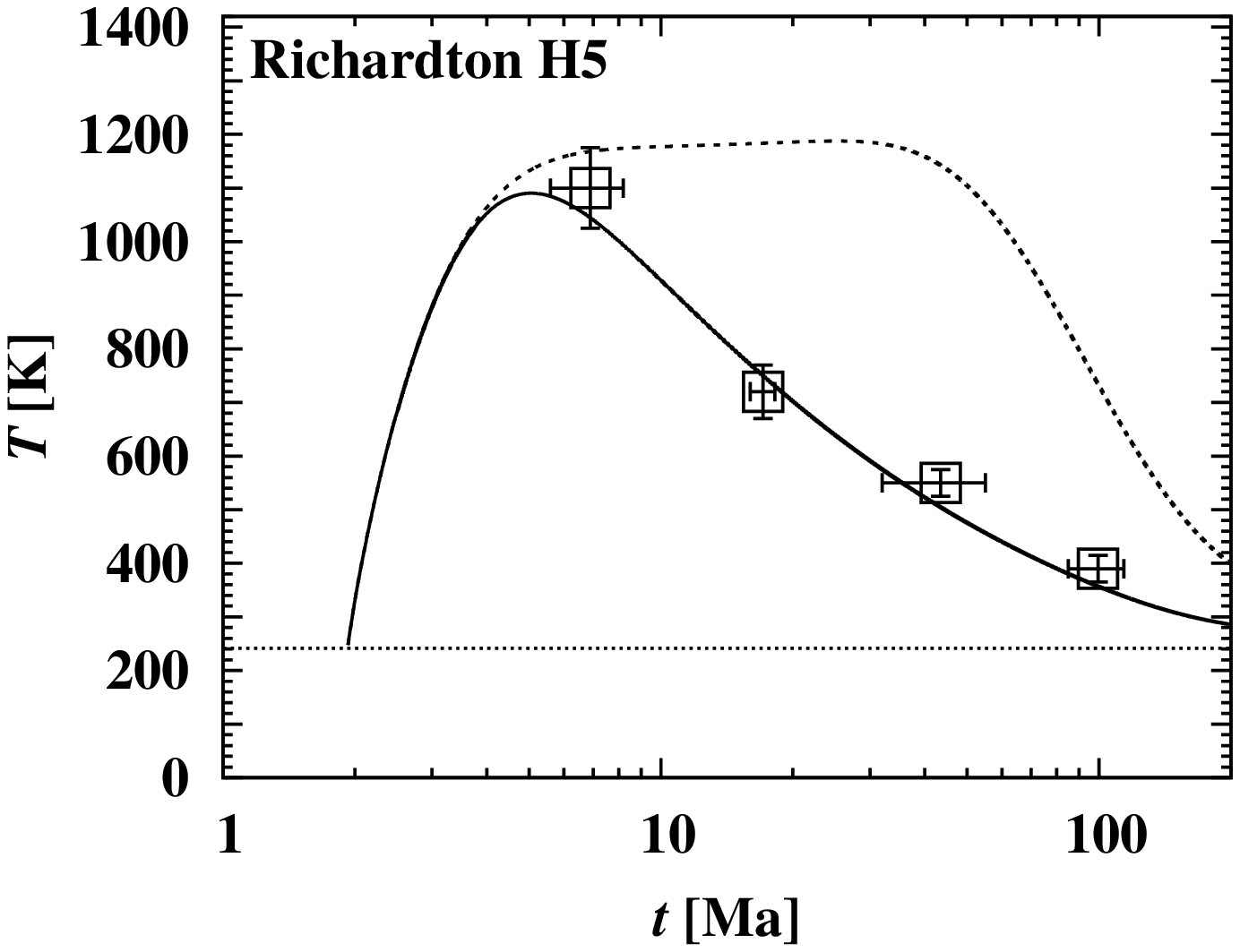}
\hfill
\includegraphics[width=0.31\hsize]{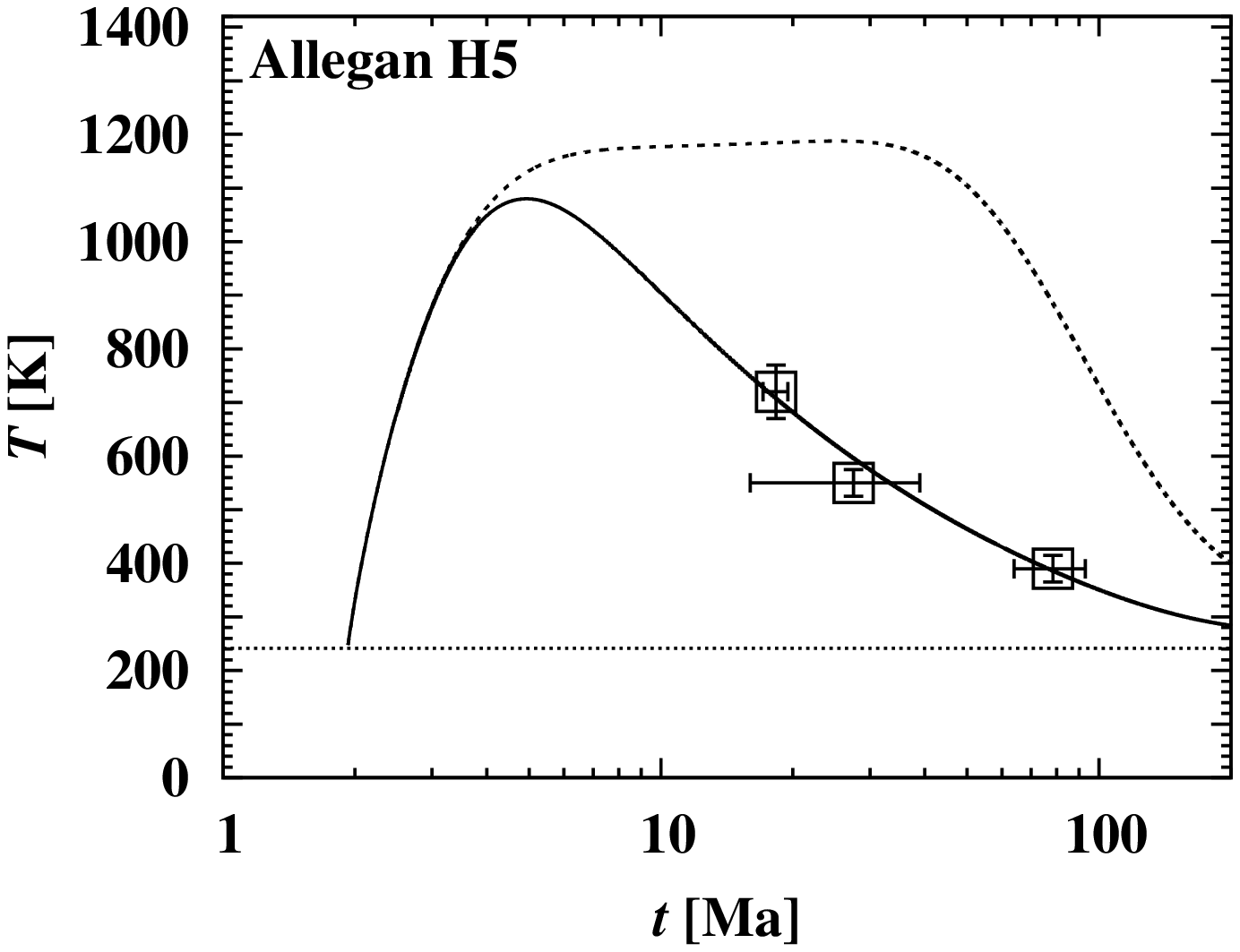} 
\hfill
\includegraphics[width=0.31\hsize]{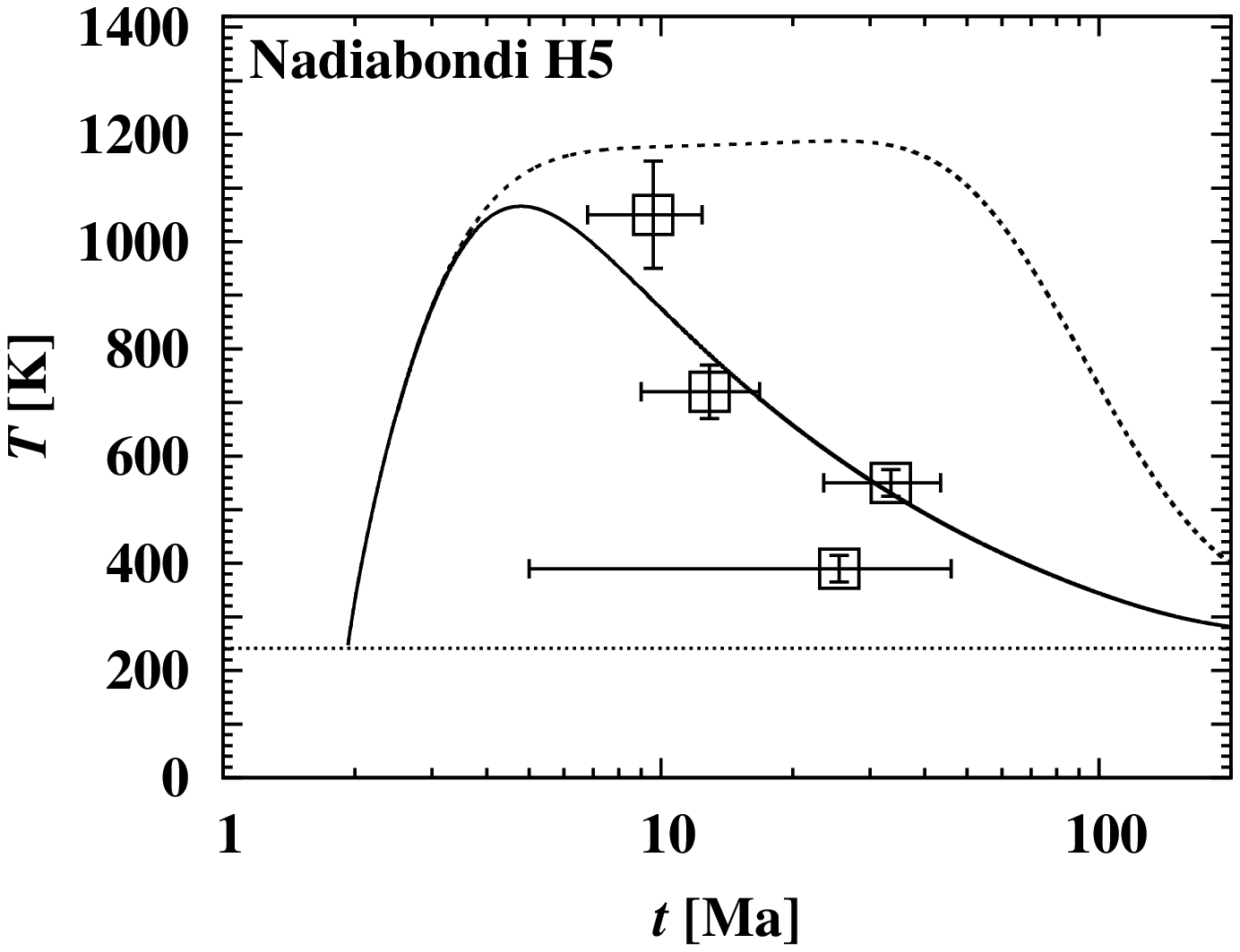} 
}

\medskip
\centerline{
\includegraphics[width=0.31\hsize]{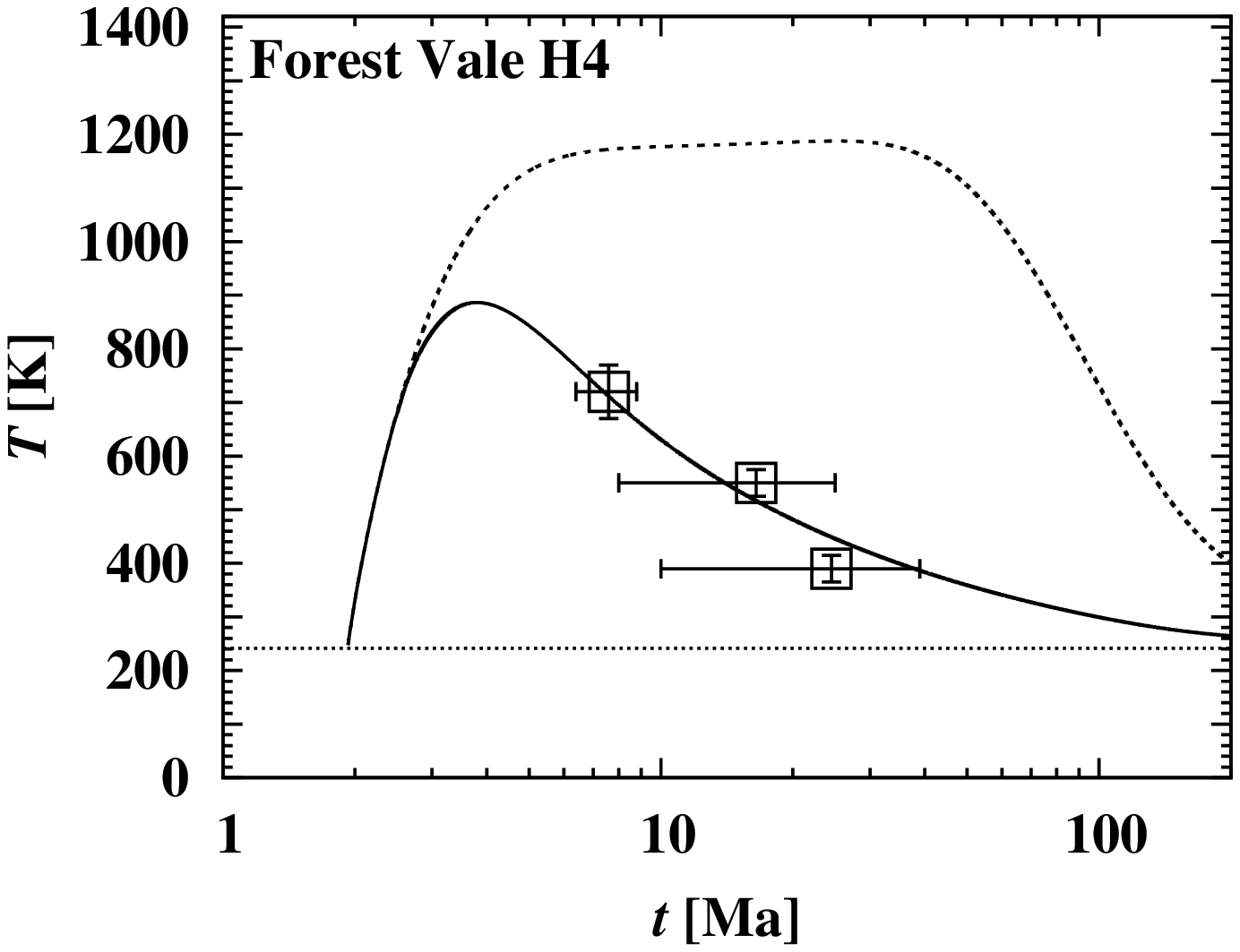}
\hspace{1em}
\includegraphics[width=0.31\hsize]{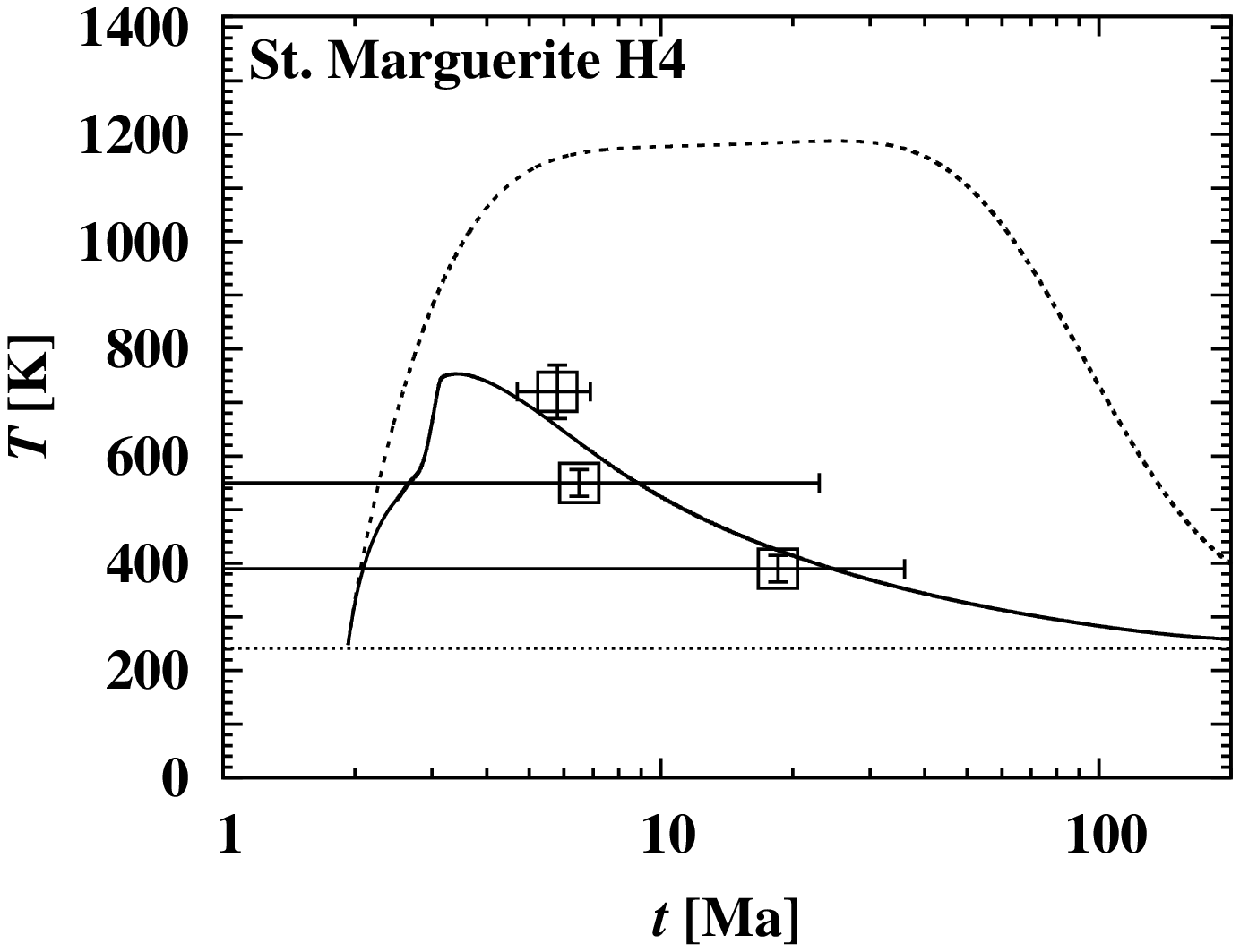}
}

\caption{Results for the individual meteorites used in the model fit. The values
of the free parameters and the burial depth's of the meteorites found for the
optimised model are given in Table~\ref{TabDepts}. The full
line shows the temperature evolution at the burial depth's of the meteorites 
in the asteroid, as determined by the evolution algorithm. The dashed line
shows the temperature evolution at the centre and the dotted 
line is the surface temperature. The  square boxes indicate the individual data
points and the error bars the uncertainty of closing temperature and age 
determinations.}
\label{S_Evo1}
\end{figure*}

\emph{Surface temperature:} This is determined by the unknown radius of the
orbit, its eccentricity, by the rotation of the body, the surface albedo, and, 
during the first few million years, by the properties of the ambient accretion
disc. All these factors are not known precisely. Therefore we think it best
to include the surface temperature $T_{\rm srf}$ as a free parameter in the
optimisation process. The resulting temperature $T_{\rm srf}$ then represents
some kind of average temperature of the surface layers during the evolution
period where the body first heats-up and then finally cools down below the 
lowest of the closure temperatures. 

The resulting value for $T_{\rm srf}$ has no bearing to the present surface
temperature of a possible successor body, because in the meantime
the body may have changed its properties by catastrophic collisions and it may
have changed to some extent its orbit during the evolution of the asteroid belt. 
Surface temperatures in the present-day asteroid belt therefore cannot be 
considered as representative for the surface temperature during the phase 
where the essential metamorphic processes acted in the body, as it is assumed in 
some thermal evolution models.

The range of surface temperatures (see Table \ref{TabVarParm}) is limited by the
fact that $T_{\rm srf}$ should be higher than the temperature where the body
would contain ice ($\approx150$\,K) and lower than the lowest of the 
closing temperatures ($\approx400$\,K). 

\emph{Surface porosity:} During the (short) main growth phase of the body the
structure of the surface layer is determined by the properties of the smaller
planetesimals that form the body under consideration. Later the surface
layer is gardened by continued impacts that produce a regolith surface layer
for which theoretical calculations indicate a porosity in the range 20\% to 30\%
\citep{Ric09,War11}. Presently, too little is known on these processes as 
that they can be included in our model calculation. Therefore we include the 
porosity of the surface layer, $\phi_{\rm srf}$, as a free parameter in the 
optimisation procedure. The resulting value then is some sort of average value 
of $\phi_{\rm srf}$ over the period where the metamorphic processes where
active.
 
\begin{table}[t]

\caption{%
Parameters and burial depths of the meteorites for the optimised H
chondrite parent body model}

\begin{tabular}{l@{\hspace{.5cm}}llll}
\hline\hline
\noalign{\smallskip}
Quantity    &  & value &  unit \\
\noalign{\smallskip}
\hline 
\noalign{\smallskip}
\multicolumn{2}{l}{Radius $R$} & 140.0 & km \\
\multicolumn{2}{l}{Formation time $t_{\rm form}$} & 1.927 & Ma \\
\multicolumn{2}{l}{Heat conductivity $K_{\rm b}$} & 4.00 & W\,m$^{-1}$K$^{-1}$ \\
\multicolumn{2}{l}{Surface temperature $T_{\rm srf}$} & 241.2 & K \\
\multicolumn{2}{l}{$^{60}$Fe/\,$^{56}$Fe ratio} &  0.0 & \\
\multicolumn{2}{l}{Surface porosity $\phi_{\rm srf}$} & 20.0\% & \\
\noalign{\smallskip}
\multicolumn{2}{l}{Maximum temperature $T_{\rm c}$} & 1\,187 & K \\
\multicolumn{1}{l}{Porous outer layer}  &              & 1.166  & km \\
\noalign{\smallskip}
\hline 
\noalign{\smallskip}
Meteorite      & type &  depth (km) & $T_{\rm max}$ (K) & $T_{\rm met}$ (K)\\
\noalign{\smallskip}
\hline 
\noalign{\smallskip}
Estacado       & H6   &   61.8           & 1174 &  $\approx1170$ \\
Guare\~na      & H6   &   60.5           & 1174 & $\approx1170$ \\
Kernouv\'e     & H6   &   48.5           & 1169 & $\approx1170$ \\
Richardton     & H5   &   18.8           & 1090 & $\approx1020$ \\
Allegan        & H5   &   17.6           & 1079 & $\approx1020$ \\
Nadiabondi     & H5   &   15.9           & 1065 & $\approx1020$ \\
Forest Vale    & H4   &  \phantom{1}5.60 & \phantom{1}886 & $\;\,\approx970$ \\
Ste. Marguerite& H4   &  \phantom{1}1.60 & \phantom{1}753 & $\;\,\approx970$ \\
\noalign{\smallskip}
\hline
\end{tabular}

\label{TabDepts}
\end{table}

\subsection{Data}

The general optimisation method of the evolution algorithm is applied to the
data for the set of H chondrites given in Table~\ref{TabDatChondCool}. 
For reasons that will become clear from the model results shown in the
subsequent figures and are discussed later the two highest closing temperature
data of Ste. Marguerite and the plutonium fission track datum of Nadiabondi are
not used. Then we have totally 28 data points that have to be fitted. The 
problem depends on the 5 parameters given in Table.~\ref{TabVarParm} and the 
eight unknown burial depths of the meteorites.

The range of surface porosities (see Table \ref{TabVarParm}) should in principle
be restricted by the requirement, that the porosity should be higher than that
of the most porous chondrites \citep[about 20\%, cf.][]{Con08}. During the model
calculations it was found that for an optimum fit values not much bigger than
this minimum value are required and, therefore, no {\it a priori} restriction
was imposed on $\phi_{\rm srf}$ during the final optimisation runs. The optimal
set of parameters, indeed, satisfies the assumption.

\subsection{Model with fixed heat conductivity}
\label{SectModelKfix}

An optimisation was run for the 5+8=13 parameter and the 28 given data points.
The model used 250 mass-shells. The optimisation was run for about 5\,500
generations with hundred individuals per generation, i.e., totally a number of
about 550\,000 complete thermal evolution models were calculated for a complete
optimisation run. This used about seven days on a normal desk-top computer
equipped with a dual-core pentium processor operating with 2.6\,GHz. 

The finally accepted optimum fit had a value of $\chi^2=5.881$ in generation 2987. This value for $\chi^2$ can be considered as excellent in view of the fact that
a good fit would require only $\chi^2\lesssim15$ (cf. Sect.~\ref{SectApplOpt}).
The model parameters of the final optimum fit for the properties of the body and
the burial depths of the meteoritic  material are given in Table~\ref{TabDepts}. 

\begin{figure*}[t]

\centerline{
\includegraphics[width=0.31\hsize]{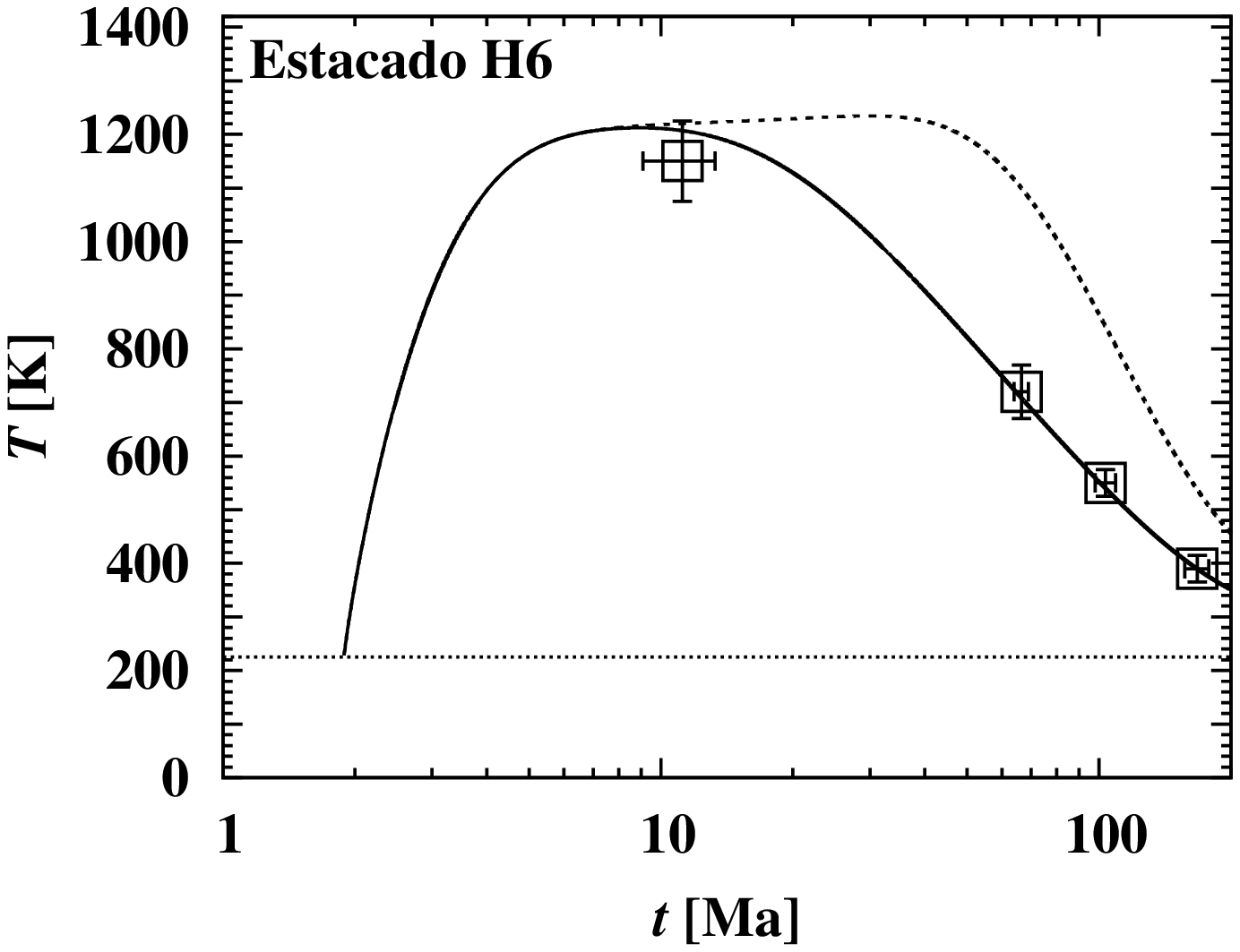}
\hfill
\includegraphics[width=0.31\hsize]{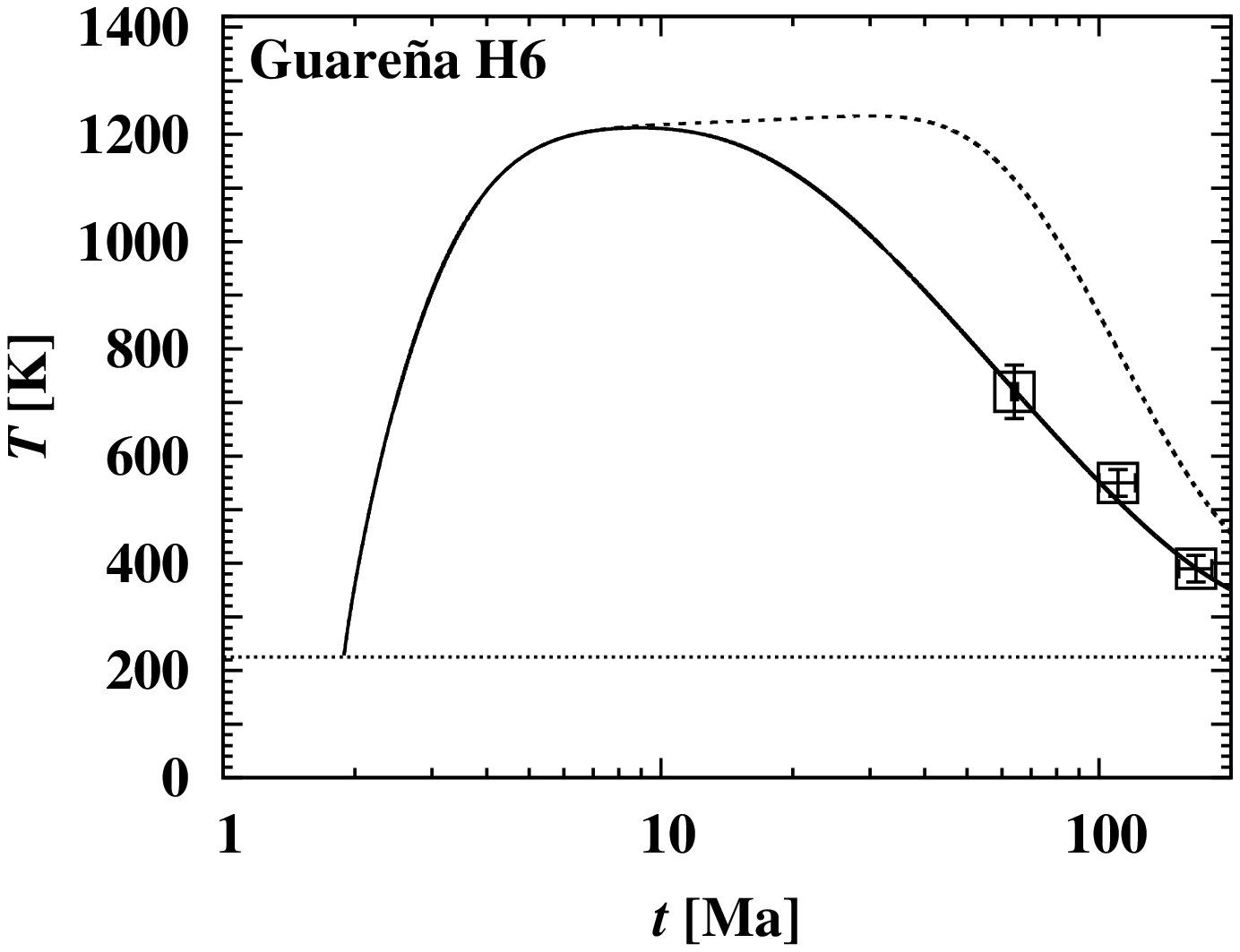}
\hfill
\includegraphics[width=0.31\hsize]{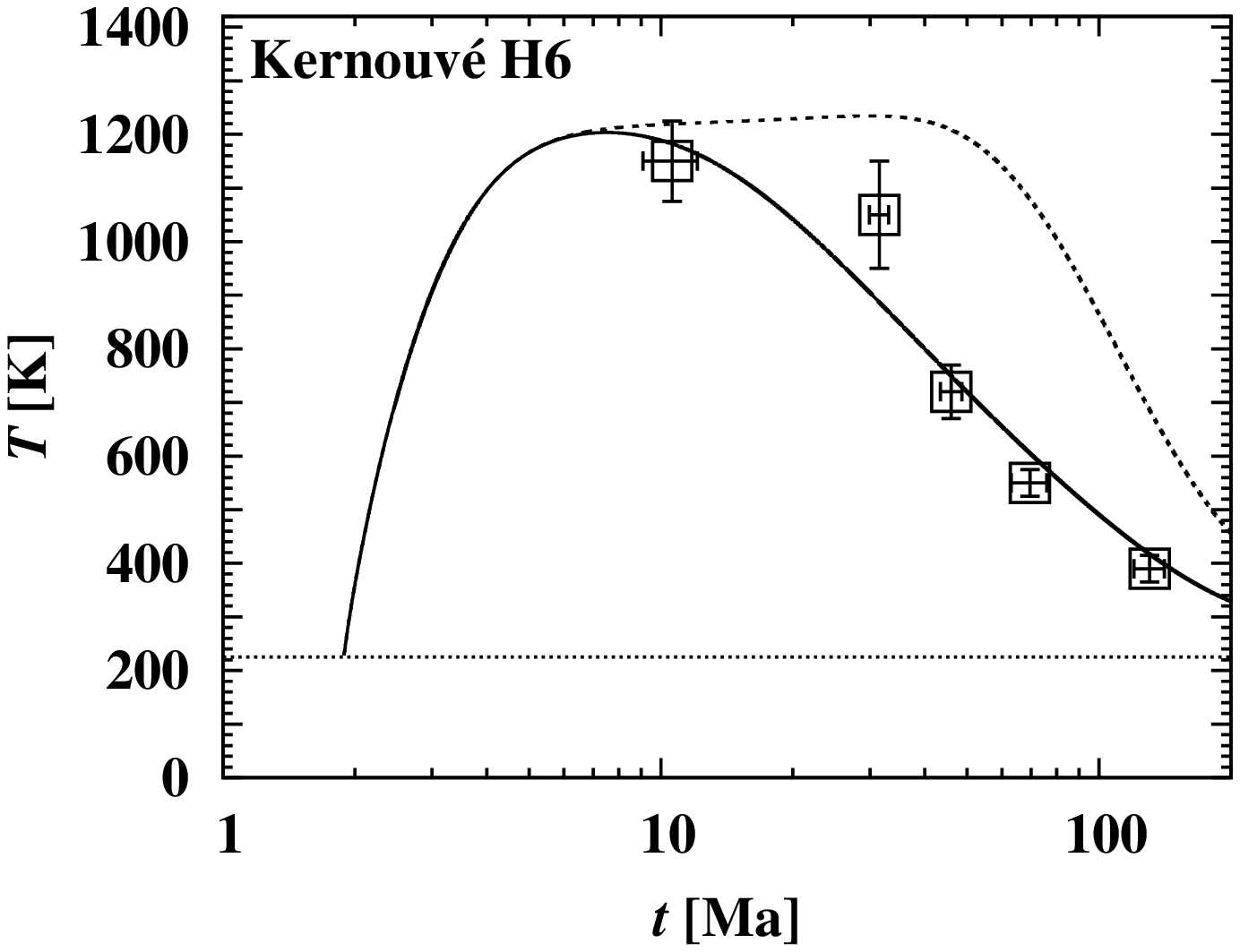} 
}

\medskip
\centerline{
\includegraphics[width=0.31\hsize]{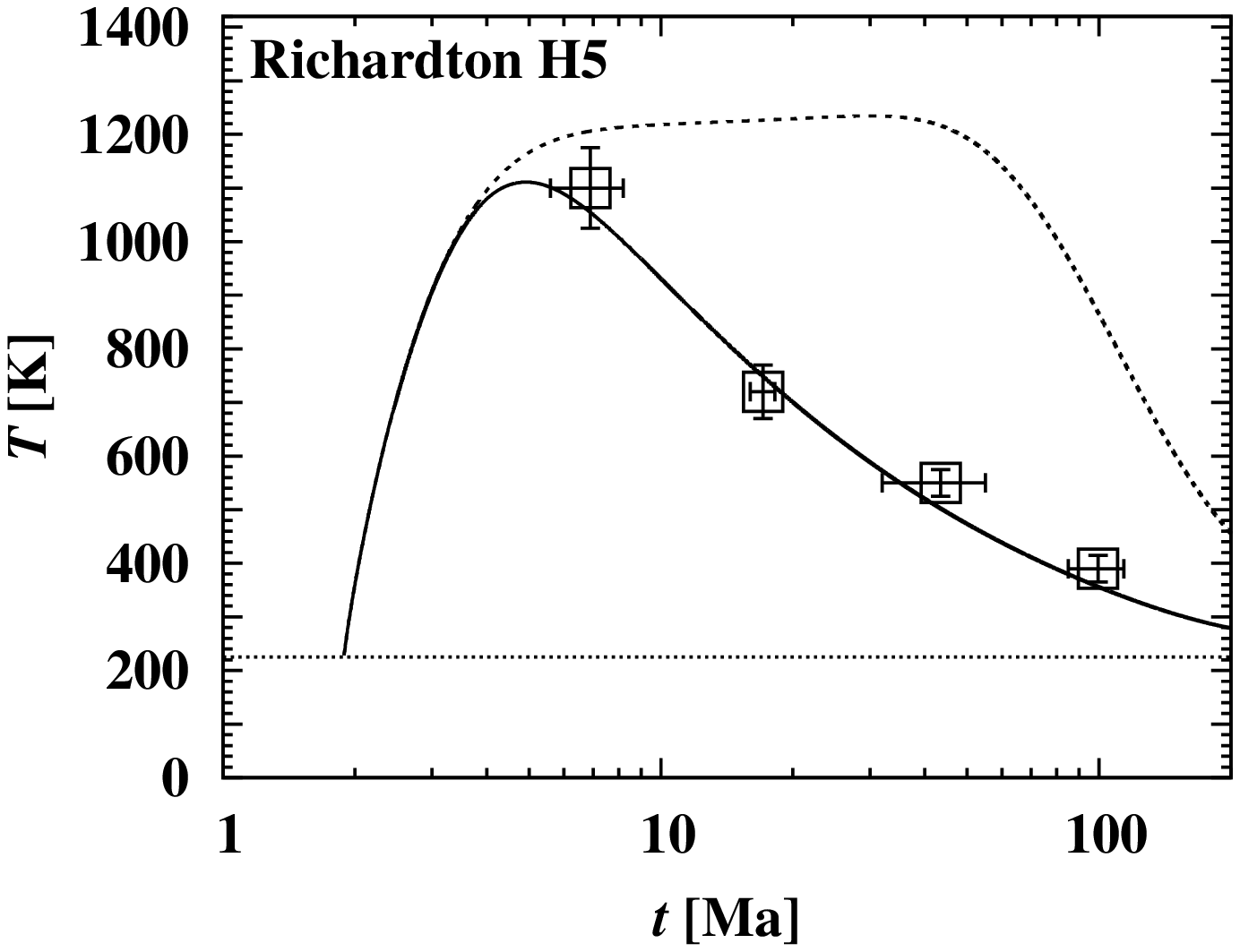}
\hfill
\includegraphics[width=0.31\hsize]{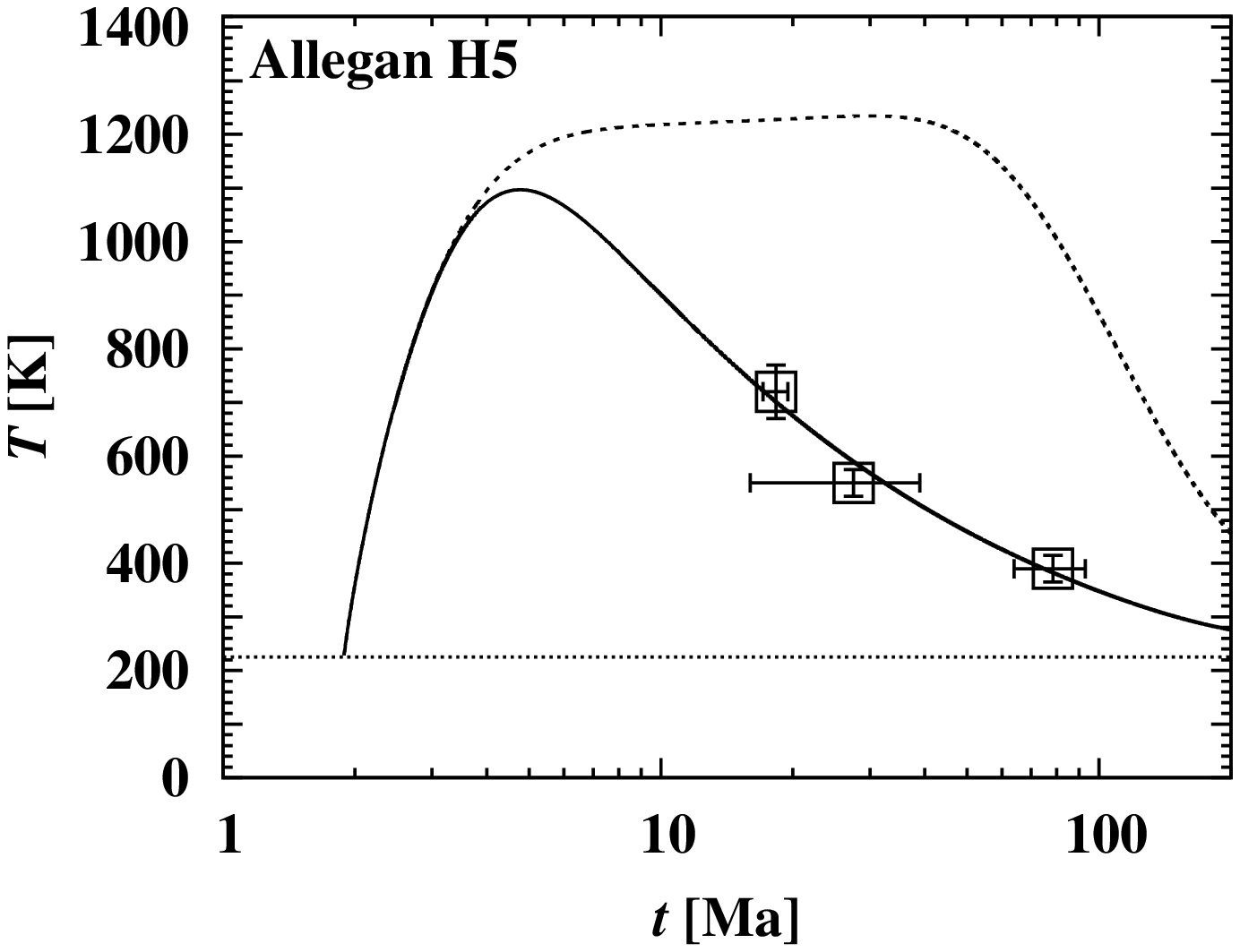} 
\hfill
\includegraphics[width=0.31\hsize]{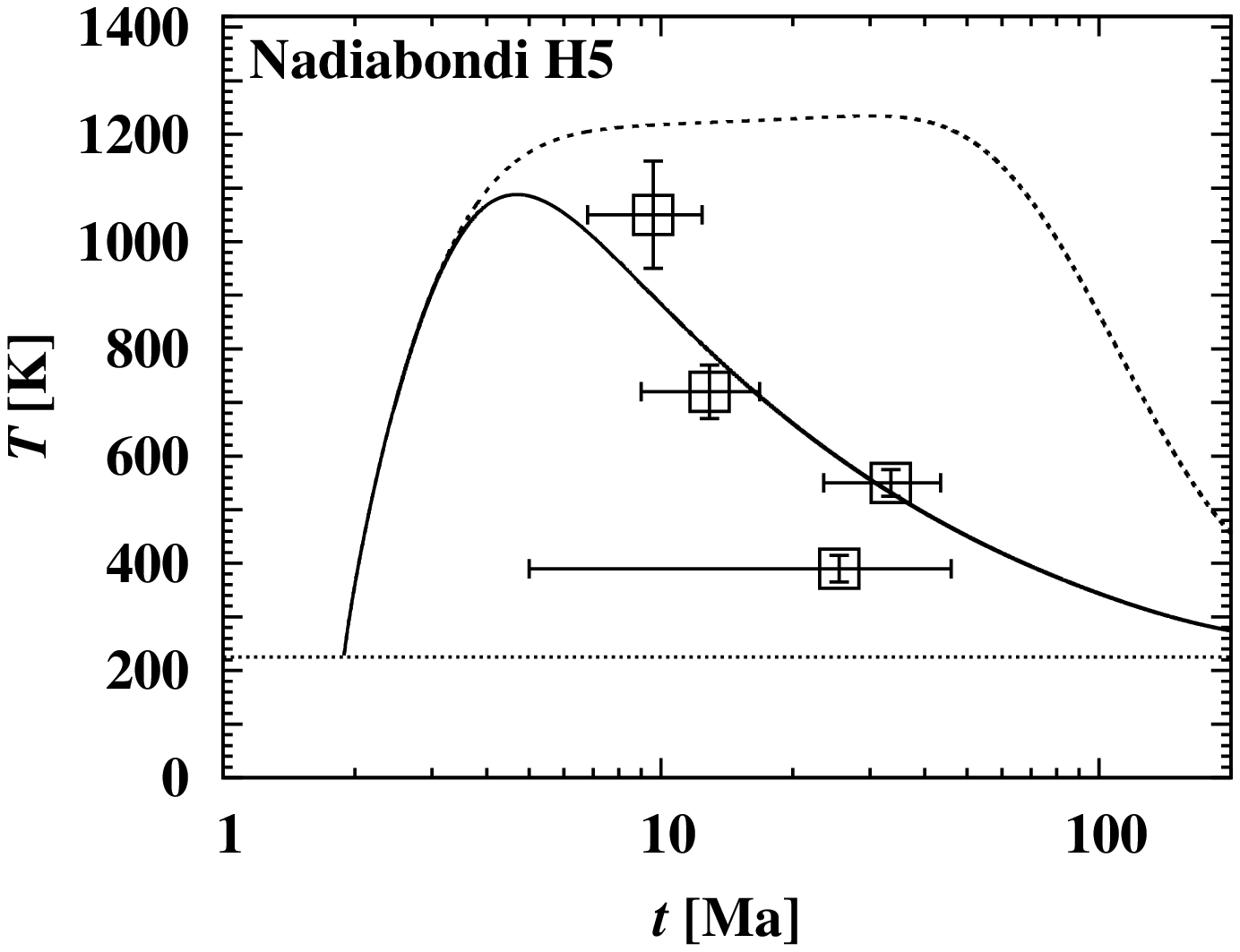} 
}

\medskip
\centerline{
\includegraphics[width=0.31\hsize]{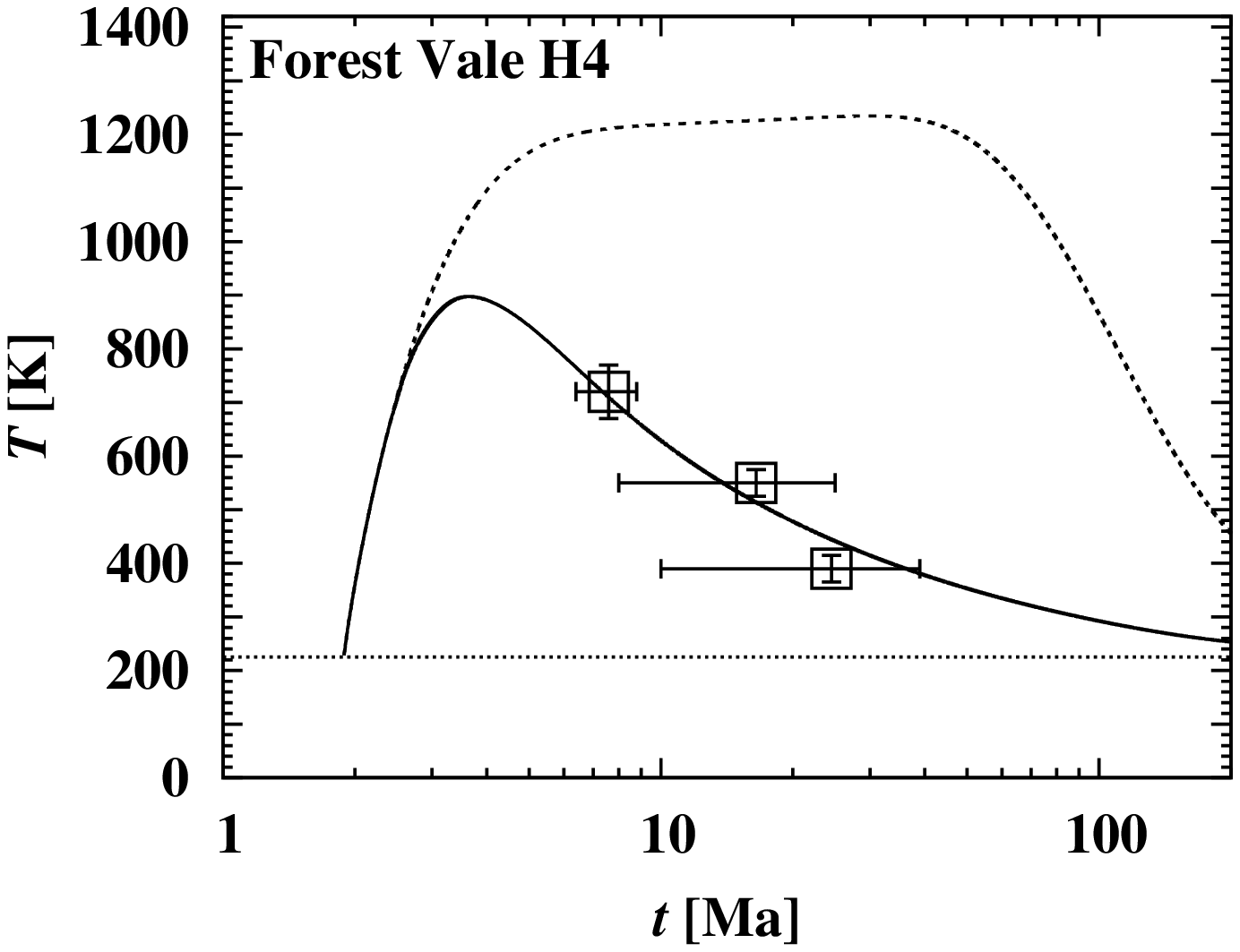}
\hspace{1em}
\includegraphics[width=0.31\hsize]{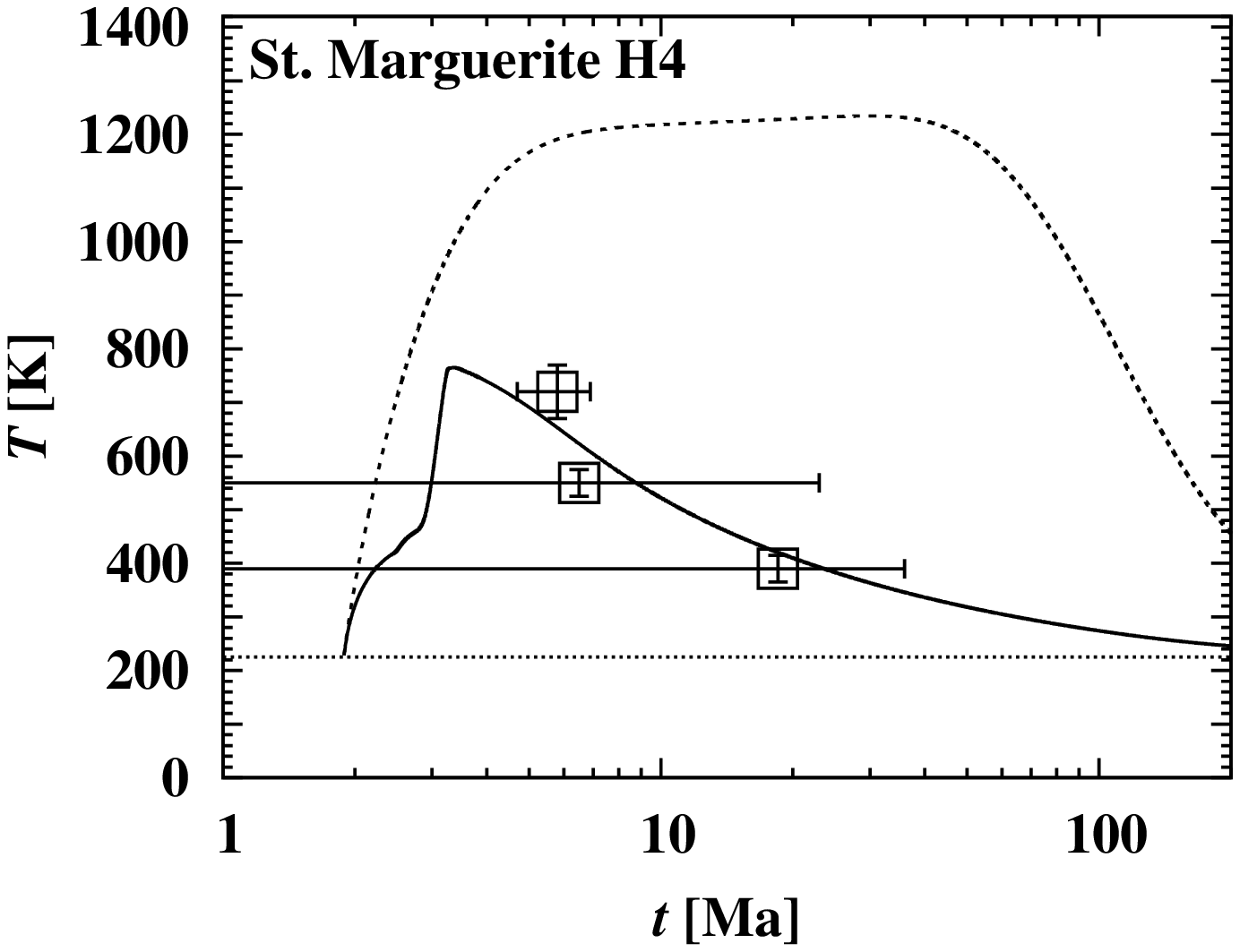}
}

\caption{Results for a model fit where the heat conductivity at zero porosity,
$K_{\rm b}$, is included in the optimisation process. The values of the free 
parameters and the burial depth's of the meteorites found for the  optimised
model are given in Table~\ref{TabDeptsVarK}. For meaning of symbols see caption
to Fig.~\ref{S_Evo1}.}
\label{FigOptVarKb}
\end{figure*}

The central temperature at maximum, $T_{\rm c}=1\,187$\,K, that results for the
set of optimum parameters is below the Fe-FeS eutectic at 1220\,K 
\citep{Fei97}. For the optimum model the temperature in the centre therefore is
below the border temperature where differentiation would just become possible. 
In so far the model is compatible with the basic assumption of the model
calculation that melting can be neglected.

\begin{table}

\caption{%
Parameter for the model with optimised $K_0$, burial depths of the meteorites,
and maximum temperature during their thermal history.}

\begin{tabular}{llllll}
\hline\hline
\noalign{\smallskip}
Quantity      & & value &  unit \\
\noalign{\smallskip}
\hline 
\noalign{\smallskip}
\multicolumn{1}{l}{Radius} & $R$ & 123.6 & km \\
\multicolumn{1}{l}{Formation time} & $t_{\rm form}$ & 1.880 & Ma \\
\multicolumn{1}{l}{Heat conductivity} & $K_{\rm b}$ & 2.65 & W\,m$^{-1}$K$^{-1}$   \\
\multicolumn{1}{l}{Surface temperature} & $T_{\rm srf}$ & 225.0 & K \\
\multicolumn{1}{l}{$^{60}$Fe/\,$^{56}$Fe ratio}& & $2.9\times10^{-7}$ & \\
\multicolumn{1}{l}{Surface porosity} & $\phi_{\rm srf}$ & 29.4\% & \\
\noalign{\smallskip}
\multicolumn{1}{l}{Maximum temperature} & $T_{\rm d}$ & 1\,234 & K \\
\multicolumn{1}{l}{Porous outer layer}  &              & 0.396 & km \\
\noalign{\smallskip}
\hline 
\noalign{\smallskip}
Meteorite      & type & \multicolumn{1}{c}{depth [km]} & 
\multicolumn{1}{c}{$T_{\rm max}$ [K]} \\
\noalign{\smallskip}
\hline 
\noalign{\smallskip}
Estacado       & H6  &   43.1              & 1\,212 \\
Guare\~na      & H6  &   43.2              & 1\,212 \\
Kernouv\'e     & H6  &   34.0              & 1\,203 \\
Richardton     & H5  &   13.7              & 1\,111 \\
Allegan        & H5  &   12.6              & 1\,096 \\
Nadiabondi     & H5  &   11.7              & 1\,087 \\
Forest Vale    & H4  &   \phantom{3}3.64   & \phantom{1}\,897 \\
Ste. Marguerite& H4  &   \phantom{3}0.638  & \phantom{1}\,764 \\
\noalign{\smallskip}
\hline
\end{tabular}

\label{TabDeptsVarK}
\end{table}

The cooling curves of the meteoritic materials at the resulting burial depths
and the empirical age-temperature data points given by the radiometric ages
are shown in Fig.~\ref{S_Evo1} for the optimised model. The solution
of the optimisation problem found here meets the data for the cooling history
of the meteorites of Table~\ref{TabDatChondCool} within their 
$1\sigma$-error-boxes in all eight cases, though in a few cases only marginally.
The fit between model and empirical data can be considered
as excellent for the three meteorites with four data points: Kernouv\'{e}, 
Richardton, and Estacado. Also for three of the meteorites with three data 
points, Guare\~na, Allegan, and Forest Vale, the model fits the data-points 
rather well. Only for Nadiabondi and Ste. Marguerite the fit is of low quality.
These two cases are discussed later.

\begin{figure*}[t]

\centerline{
\includegraphics[width=0.31\hsize]{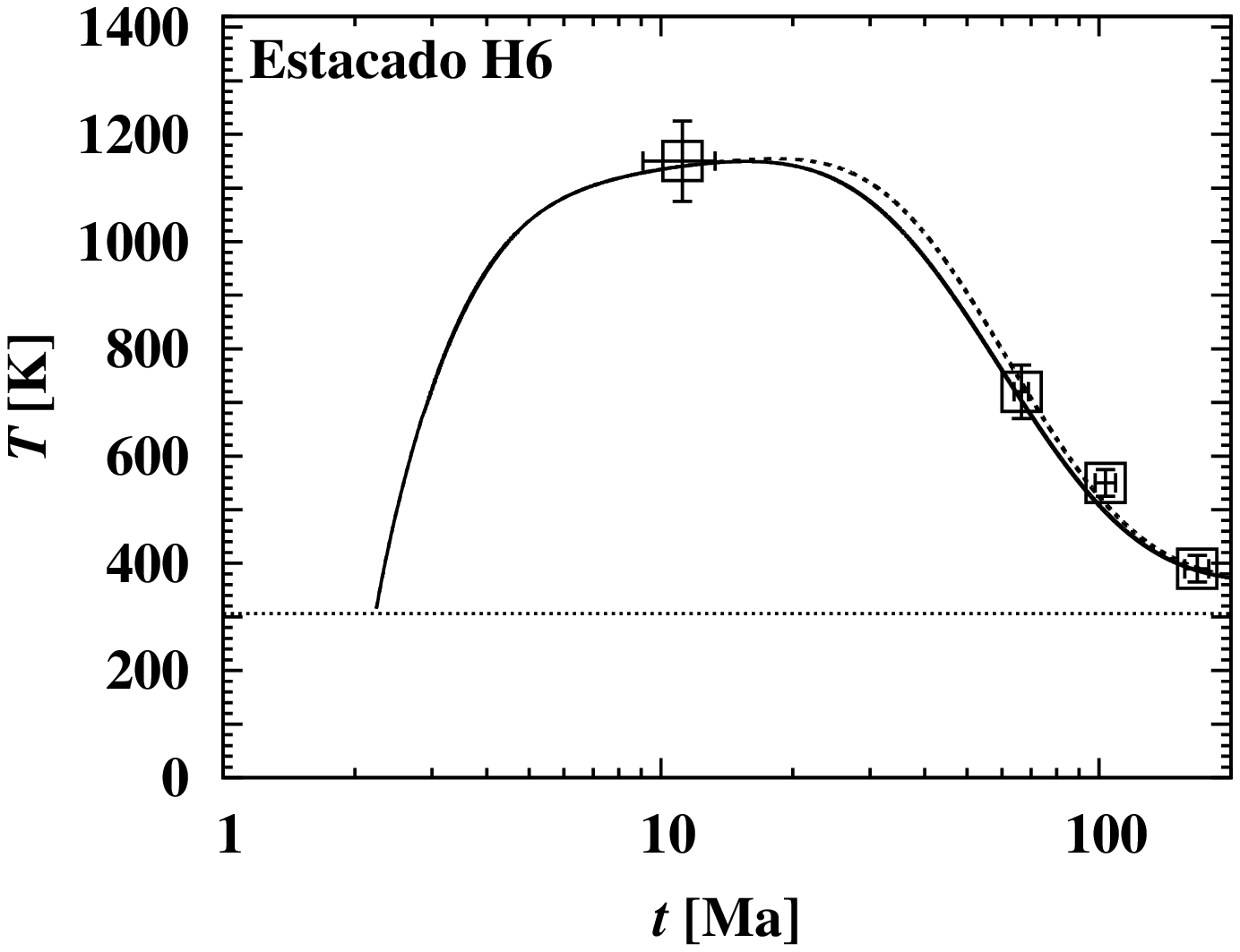}
\hfill
\includegraphics[width=0.31\hsize]{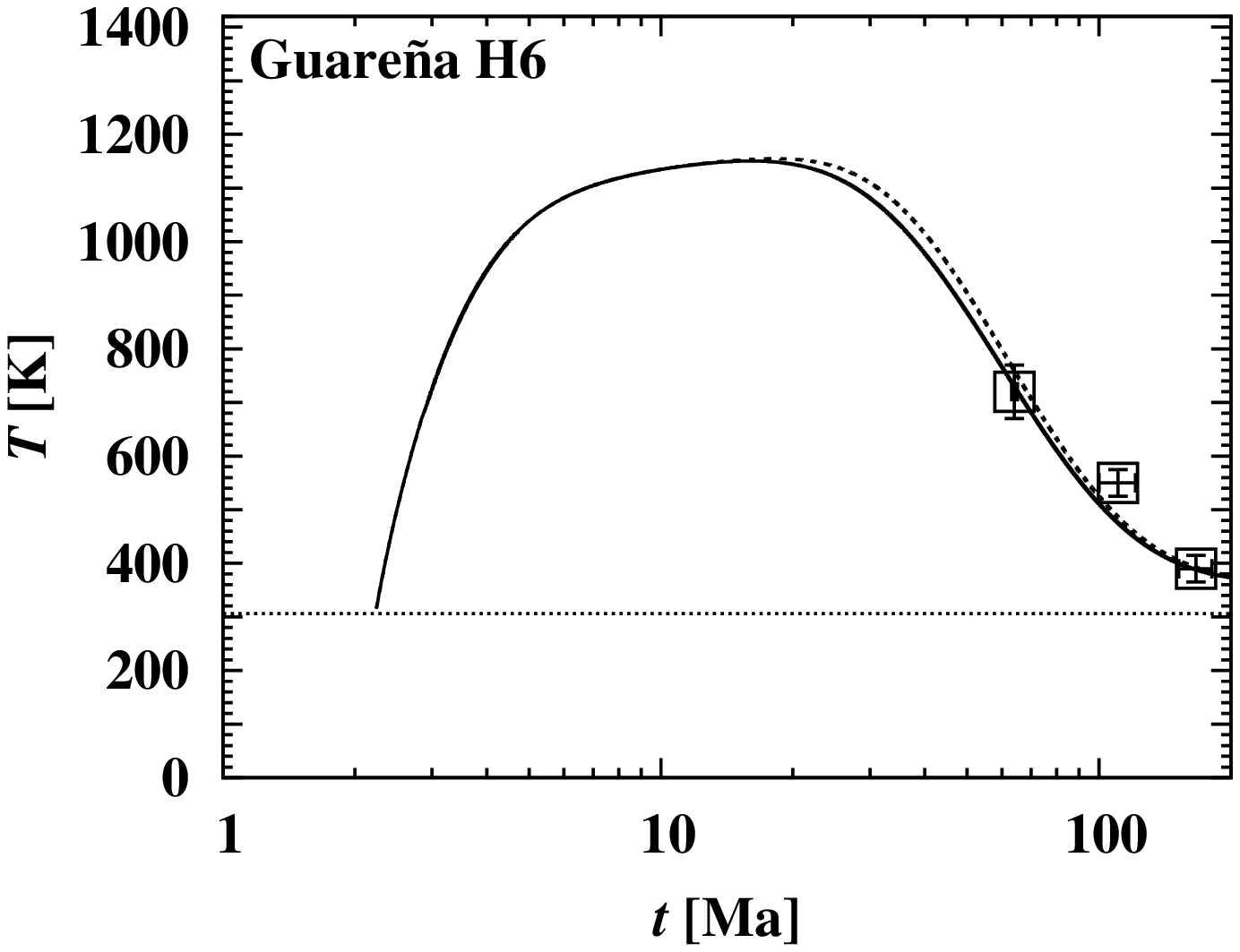}
\hfill
\includegraphics[width=0.31\hsize]{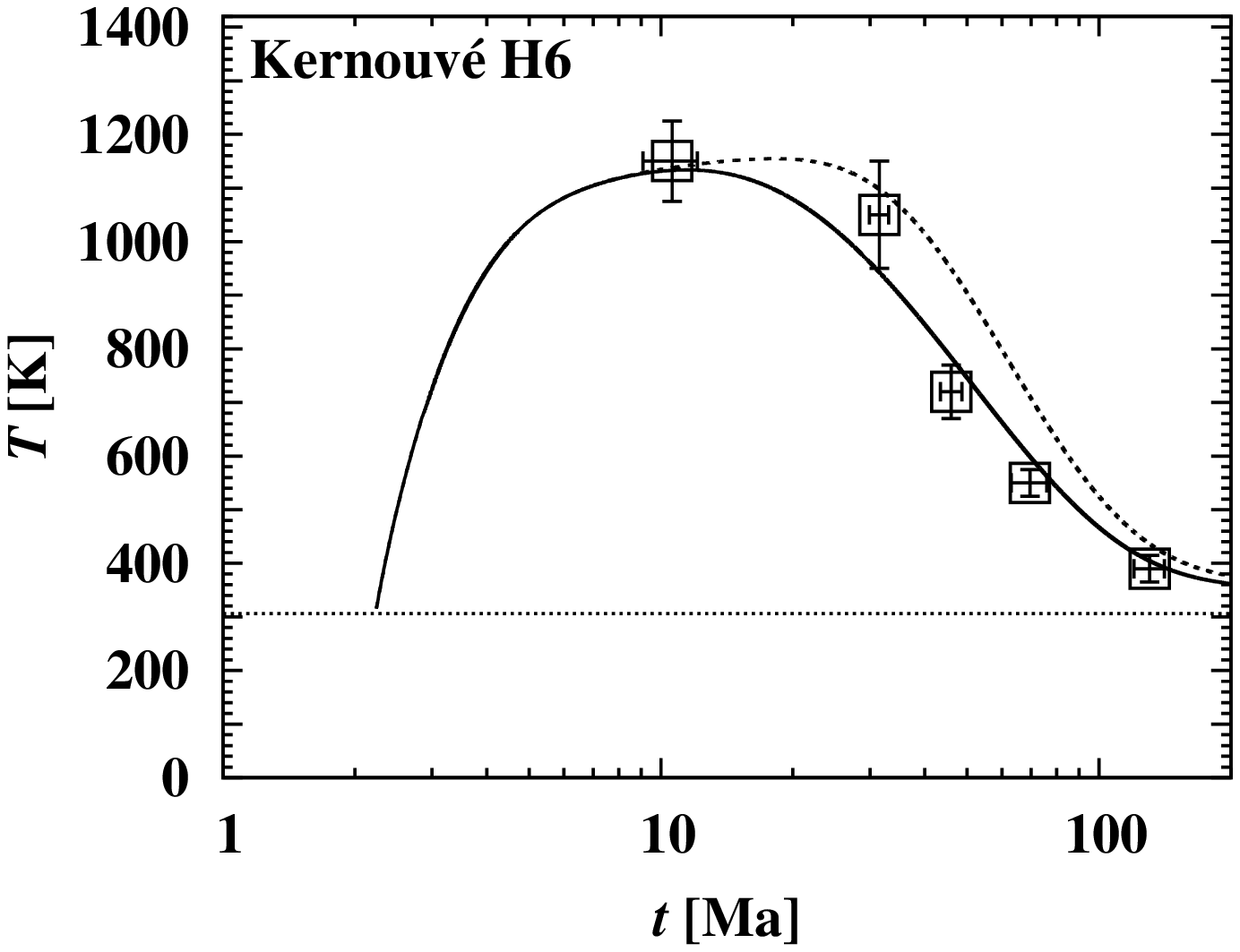} 
}

\medskip
\centerline{
\includegraphics[width=0.31\hsize]{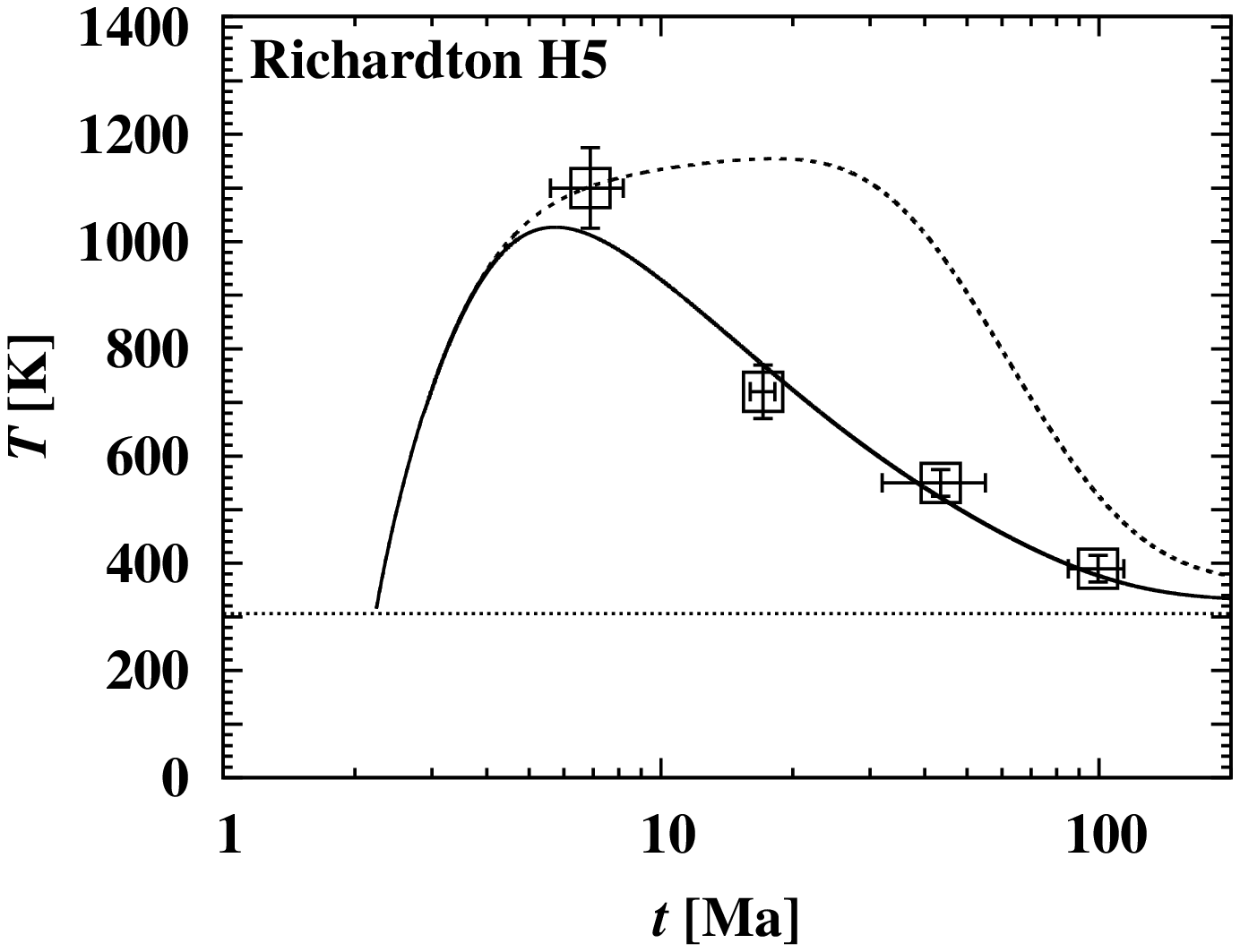}
\hfill
\includegraphics[width=0.31\hsize]{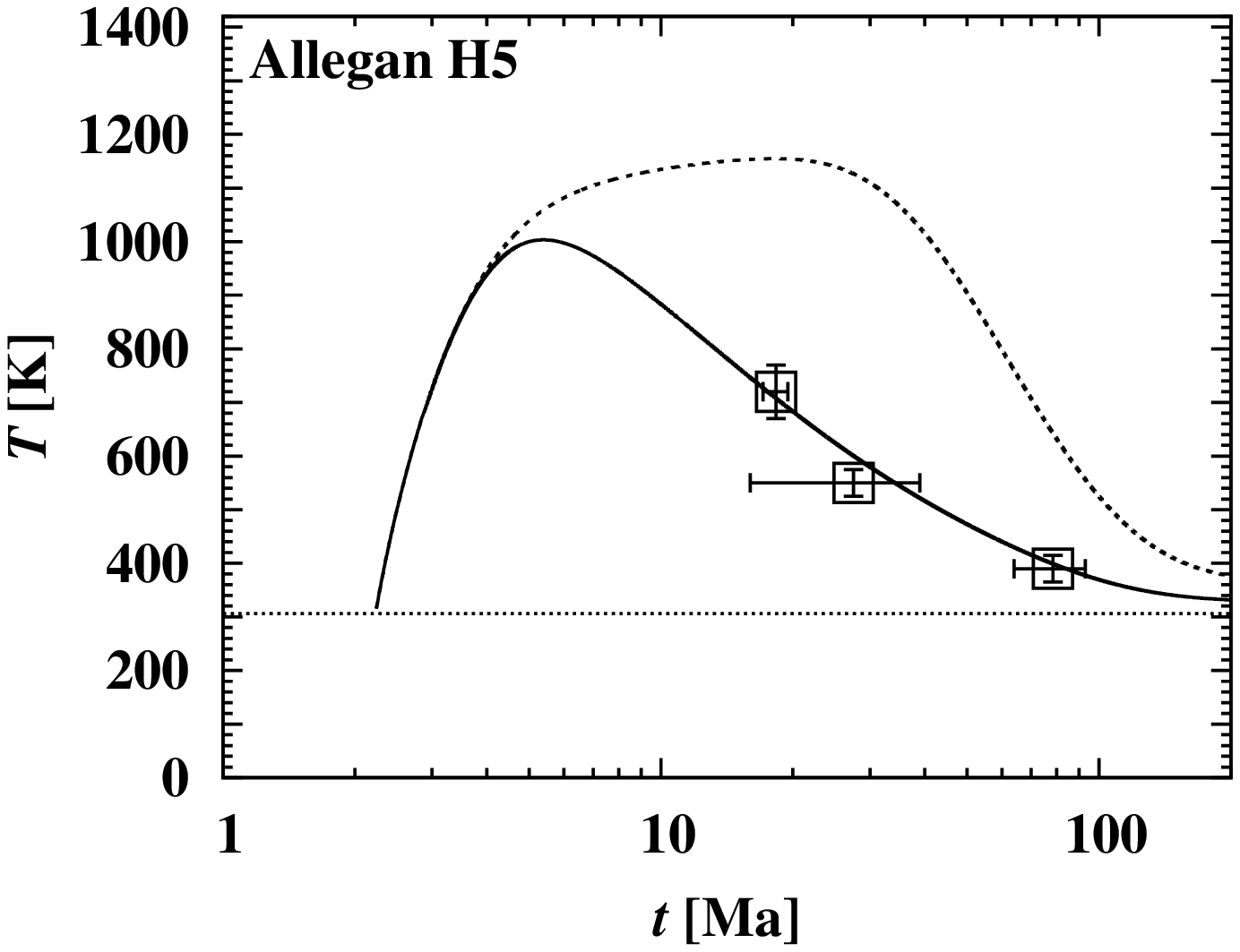} 
\hfill
\includegraphics[width=0.31\hsize]{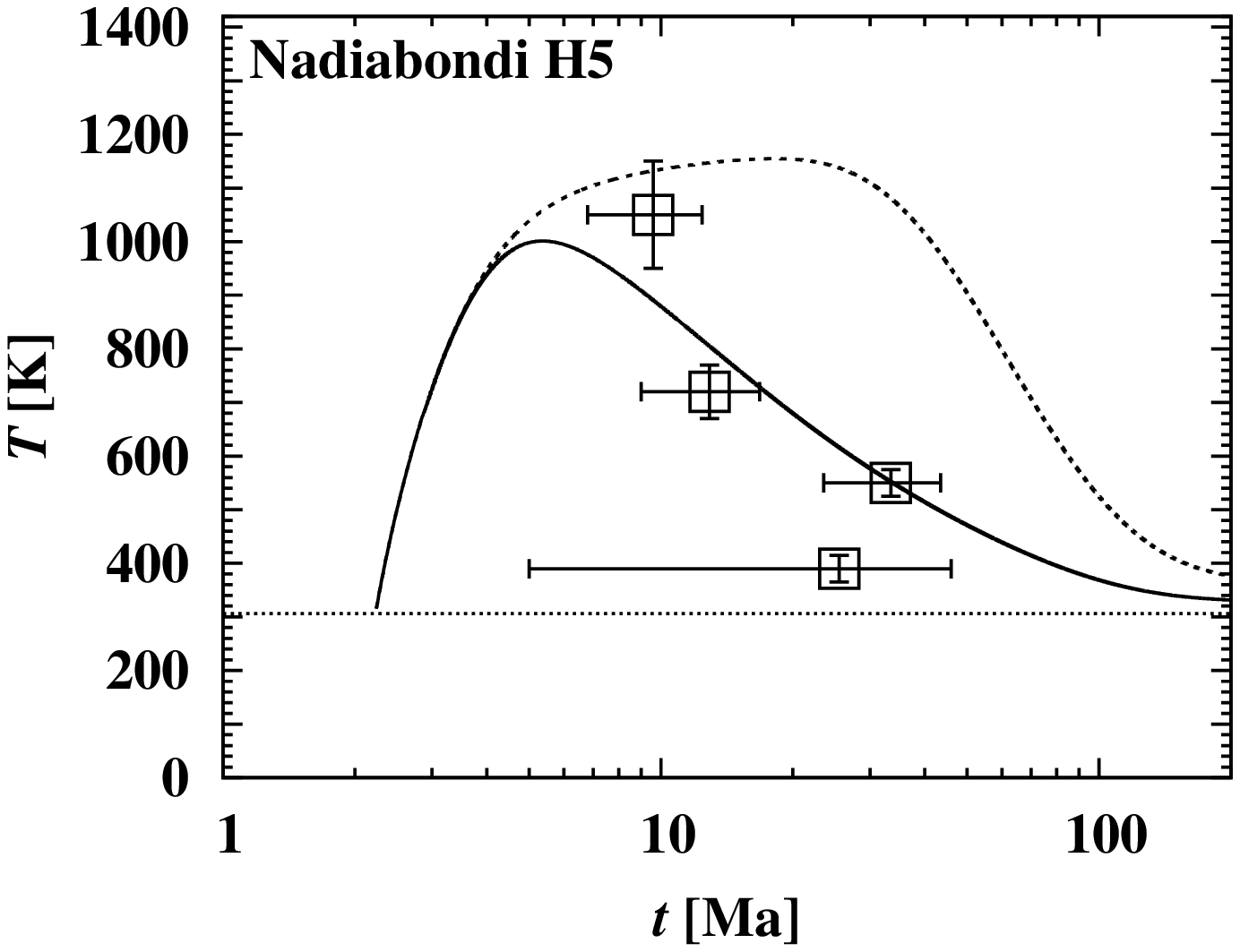} 
}

\medskip
\centerline{
\includegraphics[width=0.31\hsize]{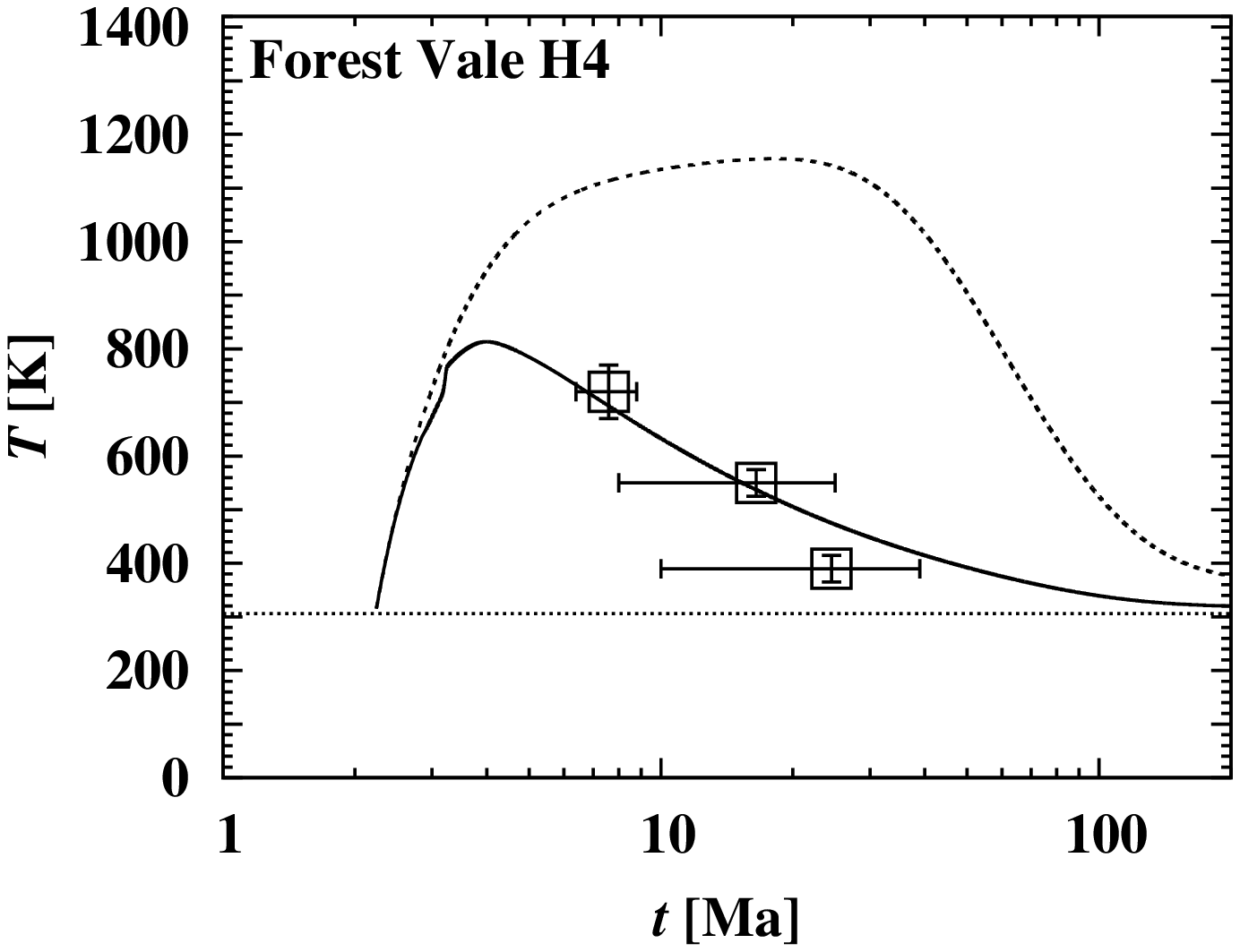}
\hspace{1em}
\includegraphics[width=0.31\hsize]{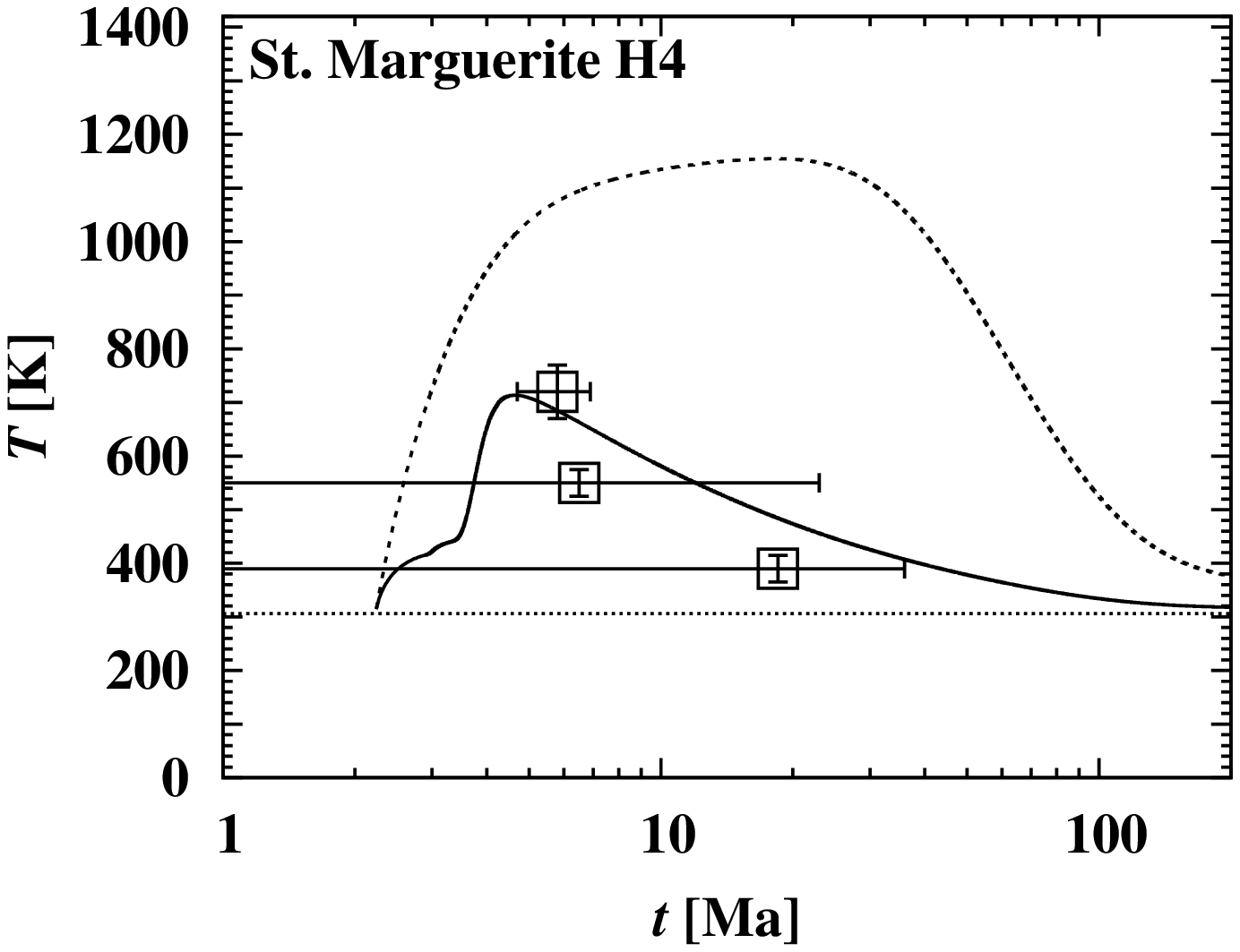}
}

\caption{Results for a model fit where the radius is set equal to the 
observed radius of (6) Hebe. The burial depth's of the meteorites found for the optimised
model are given in Table~\ref{TabDeptsHebe}. For meaning of symbols see caption 
to Fig.~\ref{S_Evo1}.}
\label{FigOptHebe}
\end{figure*}

The maximum temperature, $T_{\rm max}$, reached during the thermal history at
each of the different depths are given in Table~\ref{TabDepts}. For comparison, 
the table also gives the metamorphic temperature of the different petrologic
types, $T_{\rm met}$ \citep{Sla05}. The maximum temperatures of the different
meteorites at their burial depth, $T_{\rm max}$, fit reasonably well
with the requirement $T_{\rm max}>T_{\rm met}$. 

\begin{table}

\caption{%
Parameter for the model assuming a radius equal to that of (6) Hebe, burial 
depths of the meteorites, and maximum temperature during their thermal history.}

\begin{tabular}{llllll}
\hline\hline
\noalign{\smallskip}
Quantity      & & value &  unit \\
\noalign{\smallskip}
\hline 
\noalign{\smallskip}
\multicolumn{1}{l}{Radius} & $R$ & 93.0 & km \\
\multicolumn{1}{l}{Formation time} & $t_{\rm form}$ & 2.239 & Ma \\
\multicolumn{1}{l}{Heat conductivity} & $K_{\rm b}$ & 2.92 & W\,m$^{-1}$K$^{-1}$   \\
\multicolumn{1}{l}{Surface temperature} & $T_{\rm srf}$ & 306.5 & K \\
\multicolumn{1}{l}{$^{60}$Fe/\,$^{56}$Fe ratio}& & $2.0\times10^{-7}$ & \\
\multicolumn{1}{l}{Surface porosity} & $\phi_{\rm srf}$ & 34.0\% & \\
\noalign{\smallskip}
\multicolumn{1}{l}{Maximum temperature} & $T_{\rm c}$ & 1\,154 & K \\
\multicolumn{1}{l}{Porous outer layer}  &              & 0.525  & km \\
\noalign{\smallskip}
\hline 
\noalign{\smallskip}
Meteorite      & type & \multicolumn{1}{c}{depth [km]} & \multicolumn{1}{c}{$T_{\rm max}$ [K]} \\
\noalign{\smallskip}
\hline 
\noalign{\smallskip}
Estacado       & H6  &   69.7             & 1\,149 \\
Guare\~na      & H6  &   70.9             & 1\,150 \\
Kernouv\'e     & H6  &   47.9             & 1\,133 \\
Richardton     & H5  &   14.7             & 1\,026 \\
Allegan        & H5  &   12.2             & 1\,003 \\
Nadiabondi     & H5  &   12.0             & 1\,001 \\
Forest Vale    & H4  &   \phantom{3}1.85  & \phantom{1}\,813 \\
Ste. Marguerite& H4  &   \phantom{3}0.396 & \phantom{1}\,713 \\
\noalign{\smallskip}
\hline
\end{tabular}

\label{TabDeptsHebe}
\end{table}

The radius of 140\,km required for the optimum model is significantly larger
than the present day radius of (6) Hebe. If this asteroid is really the parent
body of the H-chondrites this would mean, that a layer with an average thickness
of nearly 50\,km has been eroded over time from the surface of (6) Hebe.
This is more than the depth of the layer where in this model the petrologic type
changes from H5 to H6. The change from initial to present radius amounts to
33\% or corresponds to a volume change of 71\%, i.e., the body would have lost 
most of its initial mass and all H3 and H4 chondrite material should
have been removed, which is inconsistent with meteoritic record. Probably our
optimised radius is too big and the most likely reason for this is that the
treatment of heat conduction in the porous medium and of the sintering process
is not yet realistic enough. The other possibility is that H-chondrites do not
come from (6) Hebe, although the arguments favouring (6) Hebe as parent body are
strong \citep{Gaf98}.

The thickness of the remaining porous outer layer is also given in 
Table~\ref{TabDepts}. As lower boundary of this layer we define the radius
where $\phi(r)=10$\%. At this degree of porosity usually the connections between
pores start to close and the pore structure changes from a network of
interconnected pores to isolated voids.

The low value of the surface porosity $\phi_{\rm srf}=0.20$ may either result
from the fact, that planetesimals of the 100\,km size-class gain most of their 
mass by collisions with smaller planetesimals that have already diameters of
about 10\,km and more and, therefore, are already largely compacted (see 
Fig.~10 of paper I) if they are incorporated into the growing larger body. A 
second mechanism that may be responsible for the low value of $\phi_{\rm srf}$ 
is that our optimized value represents a mean value of the actual value of 
$\phi_{\rm srf}$ over the initial heating and the extended subsequent cooling
period of 100\,Ma. During that period impacts certainly continue with a
decreasing rate that do not contribute to a significant further growth, but that
form a regolith surface. According to model calculations this has a saturation
porosity of 20\% to 30\% \citep{Ric09,War11} that would fit to the value of
$\phi_{\rm srf}=0.20$ found by the optimisation process.
 
\begin{figure}

\centerline{
\includegraphics[width=.99\hsize]{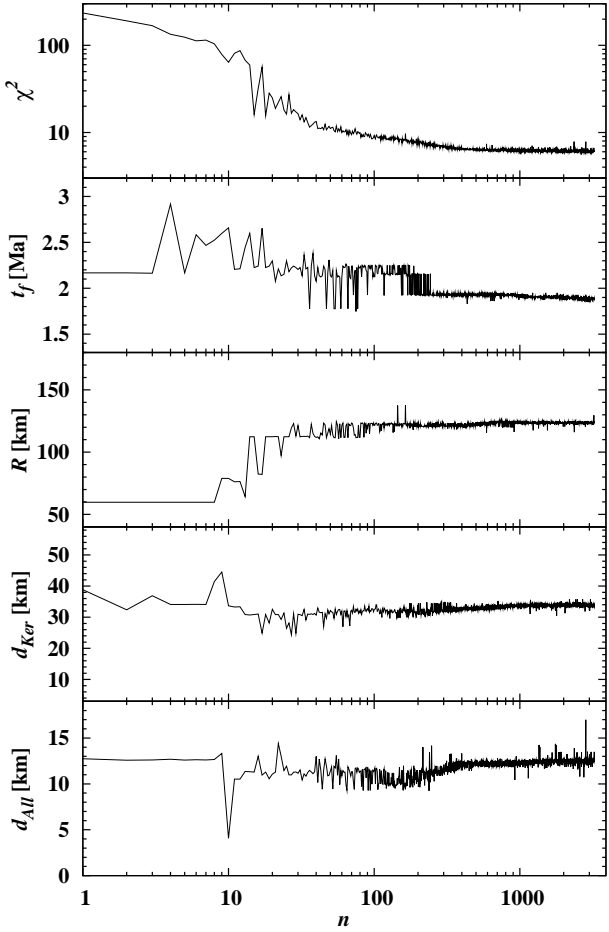}
}

\caption{Evolution history of the value of $\chi^ 2$ during the optimisation
process with progressing number of generations, $n$, for the model where the
heat conductivity at zero porosity, $K_{\rm b}$, is included in the
optimisation process. Shown is for each generation the particular individual
with the lowest value of $\chi^2$ of all individuals of that generation,
the formation time, $t_{\rm form}$, of this individual, and the radius of the 
body, $R$. 
For two representative meteorites, Kernouv\'e and Allegan, the evolution of the optimum value of the burial depth, $d$, is also shown.
}

\label{FigVarChi2}
\end{figure}

\subsection{Model with optimised heat conductivity}

\begin{figure}

\includegraphics[width=\hsize]{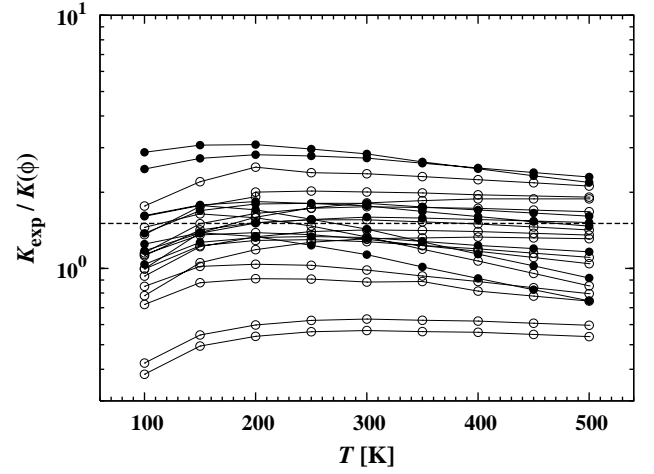}

\caption{Data for thermal conductivity $K_{\rm exp}$ of ordinary chondrites 
\citep[from][]{Yom83} divided by the analytical fit $K(\phi)$ given by 
Eq. (\ref{HeatCond}) using a value of $K_{\rm b}=2.652$ obtained as optimum value
for fitting the cooling history. Full black circles represent data for H 
chondrites, open circles data for L chondrites. The dashed line corresponds
to the value of $K=4$ used in model with fixed $K$.
}

\label{FigVarKb}
\end{figure}

The pre-factor $K_{\rm b}$ in our approximation for the porosity-dependent
heat conductivity, Eq.~(\ref{HeatCond}), corresponds to the heat conductivity of
the bulk  material (i.e., at $\phi=0$ or $D=1$), determined by extrapolating
data obtained from a couple of ordinary chondrites \citep{Yom83} to zero porosity. Figure 6 in that paper and our Fig.~\ref{FigVarKb} shows that there is
considerable scatter in the data which results in some  uncertainty of the value
of $K_{\rm b}$. The value for $K_{\rm b}$ used in the above model calculation 
(see Table~\ref{TabDepts}) may therefore be an unfavourable choice from the 
allowed range of values compatible with experimental values. Additionally, there
is the problem that the small suite of chondrites used for determining 
$K_{\rm b}$ may not be representative for the bulk of the material of the parent
body. 

Therefore we performed a model calculation where $K_{\rm b}$ was considered to be a free parameter of the problem for which an optimal value is determined by the optimisation procedure. The remaining number of data-points for closure
times  with which to compare is still big enough to allow for an additional
parameter, because we now have 28 data points and 14 parameters (6 for the body
and the 8 burial depth's).

The model, again, used 250 mass-shells. The optimisation was run for about 
3\,244 generations with hundred individuals per generation. The finally accepted optimum fit had a value of $\chi^2=6.033$ in generation 3229. With
this value for $\chi^ 2$ the fit can also be considered as an excellent one in
view of the fact that an acceptable fit would require only $\chi^2\lesssim14$
(cf. Sect.~\ref{SectApplOpt}). How the value of $\chi^2$ evolved during the optimisation process for this model is shown in Fig.~\ref{FigVarChi2}.
The somewhat ``noisy'' appearance of the evolution history reflects the permanent  mutations and crossing of genes in the evolution algorithm that not always results in children that are as fit as their parents.

The model parameters of the final optimum fit for the properties of the body
and  the burial depth's of the meteoritic material are given in 
Table~\ref{TabDepts} and the model results for the individual meteorites are
shown in  Fig.~\ref{FigOptVarKb}. The already good fit to the data points shown
in  Fig.~\ref{S_Evo1} is once more slightly improved. The two meteorites
Nadiabondi and Ste. Marguerite, however, drop out again and are not well fitted.

\begin{figure*}[t]

\centerline{
\includegraphics[width=0.31\hsize]{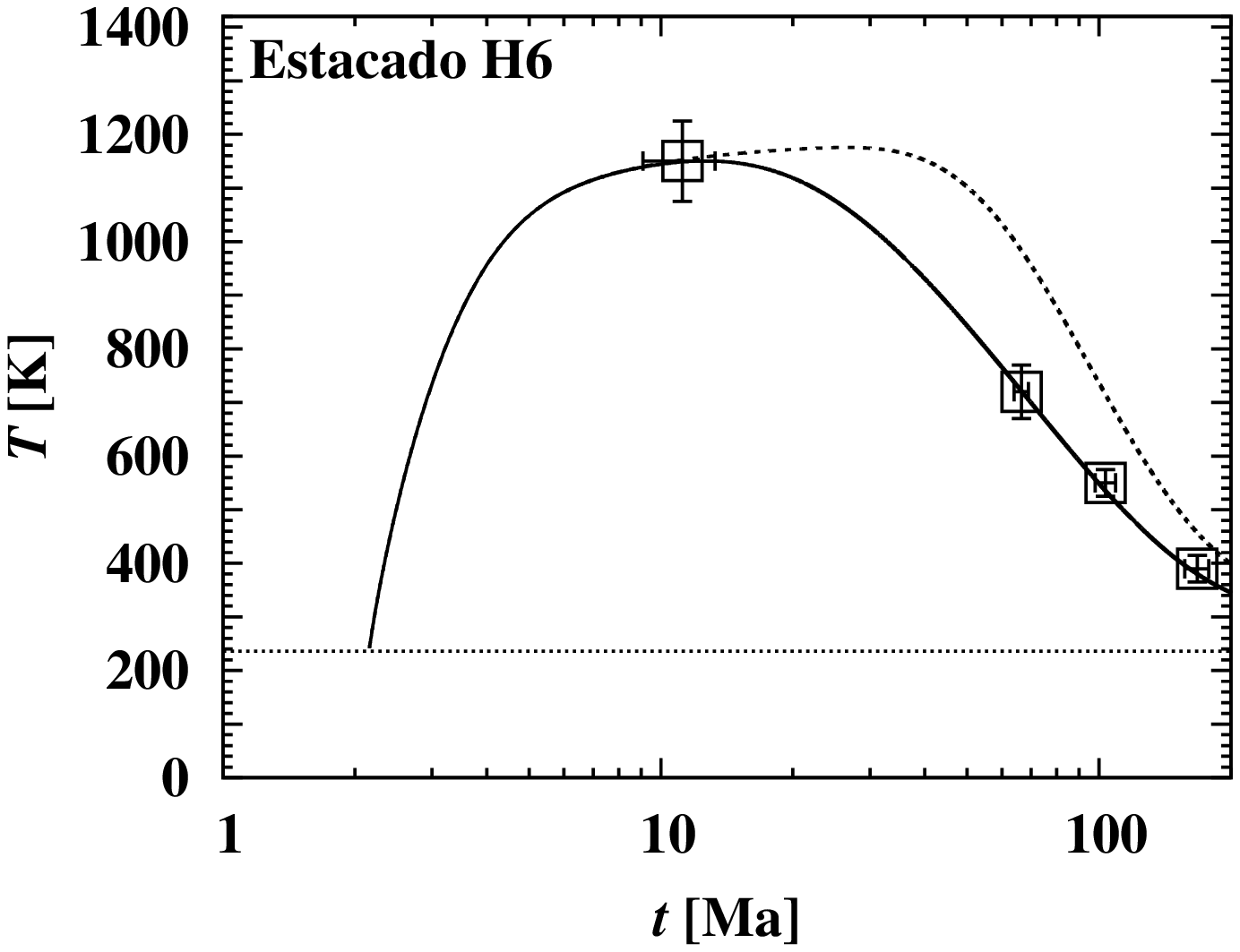}
\hfill
\includegraphics[width=0.31\hsize]{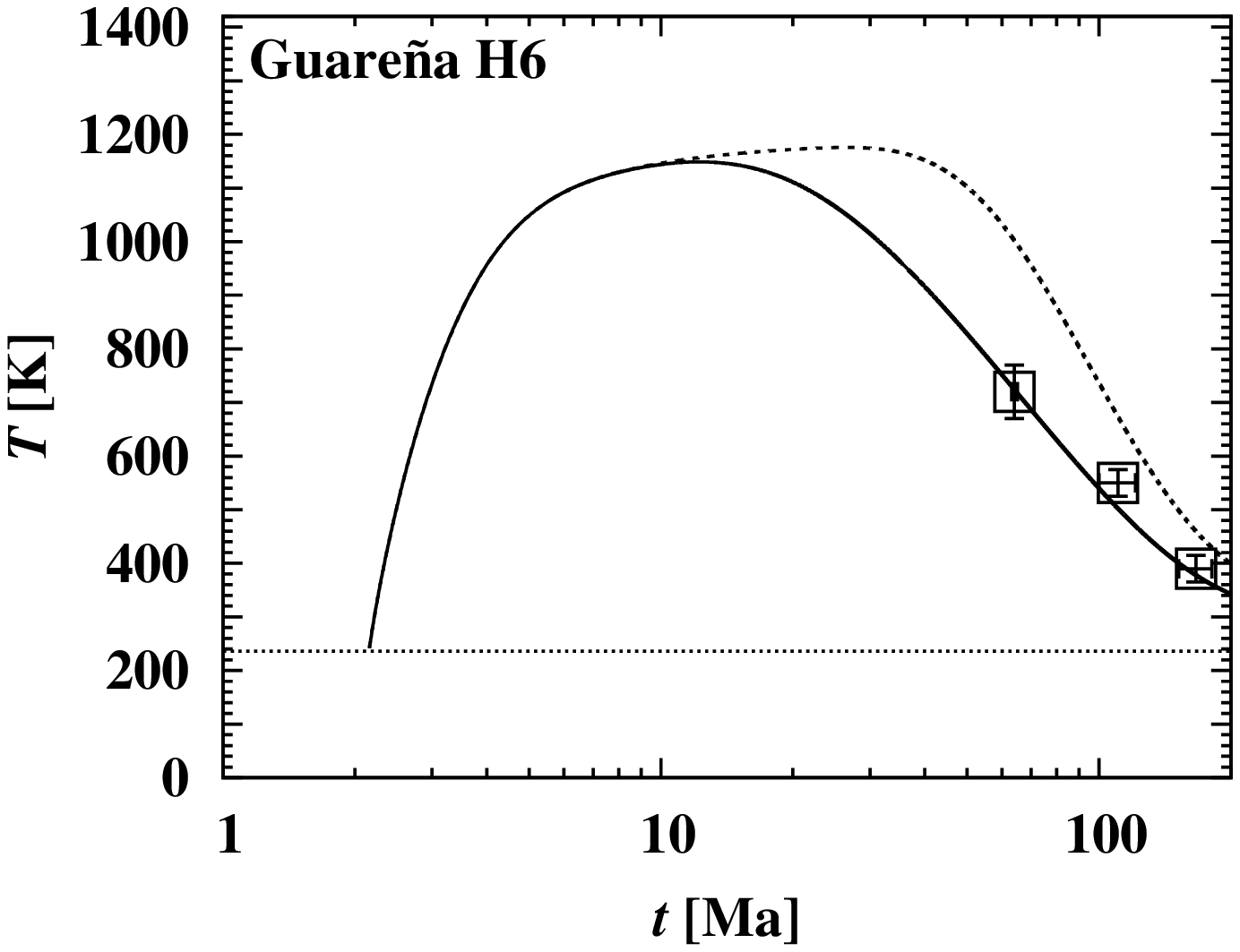}
\hfill
\includegraphics[width=0.31\hsize]{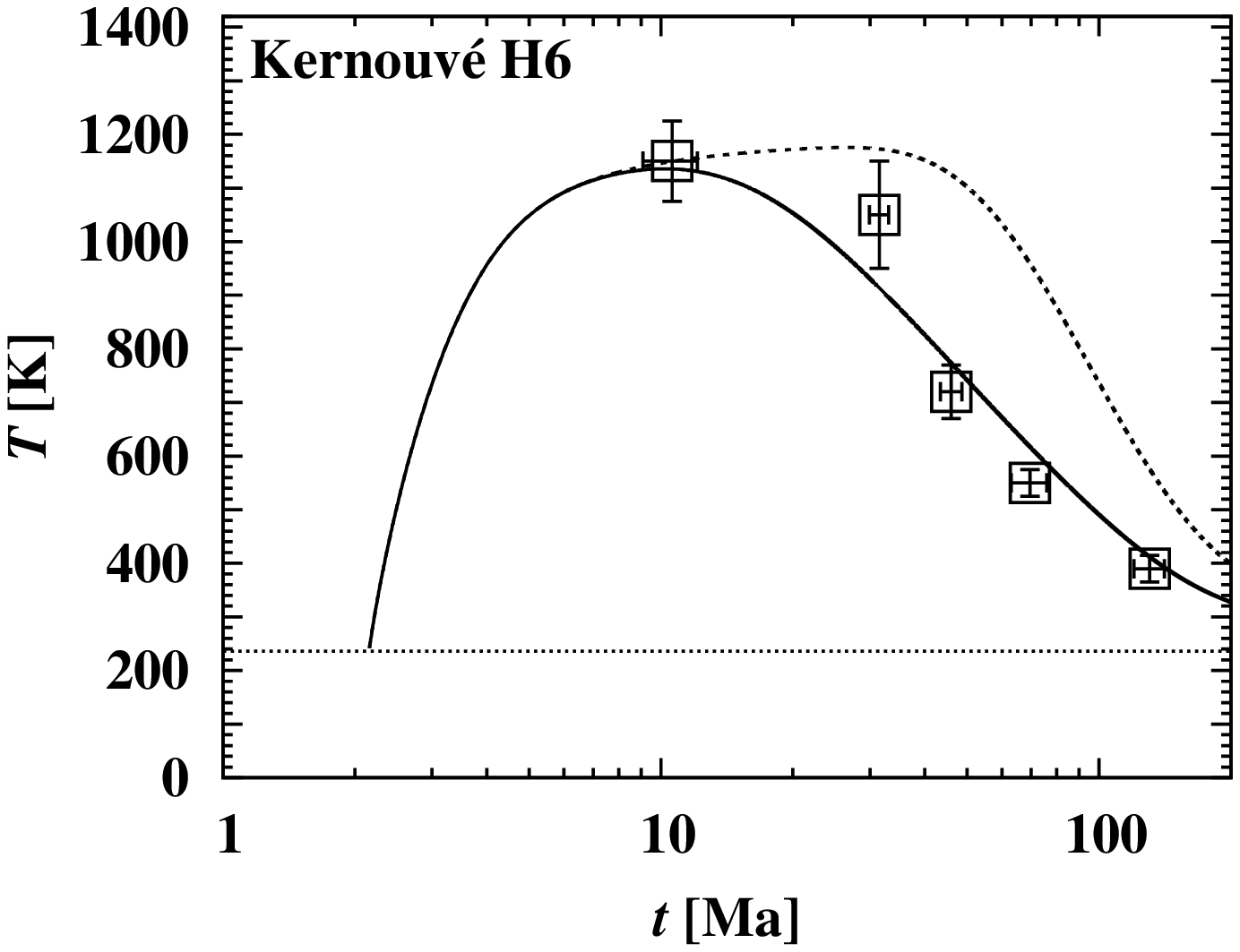} 
}

\centerline{
\includegraphics[width=0.31\hsize]{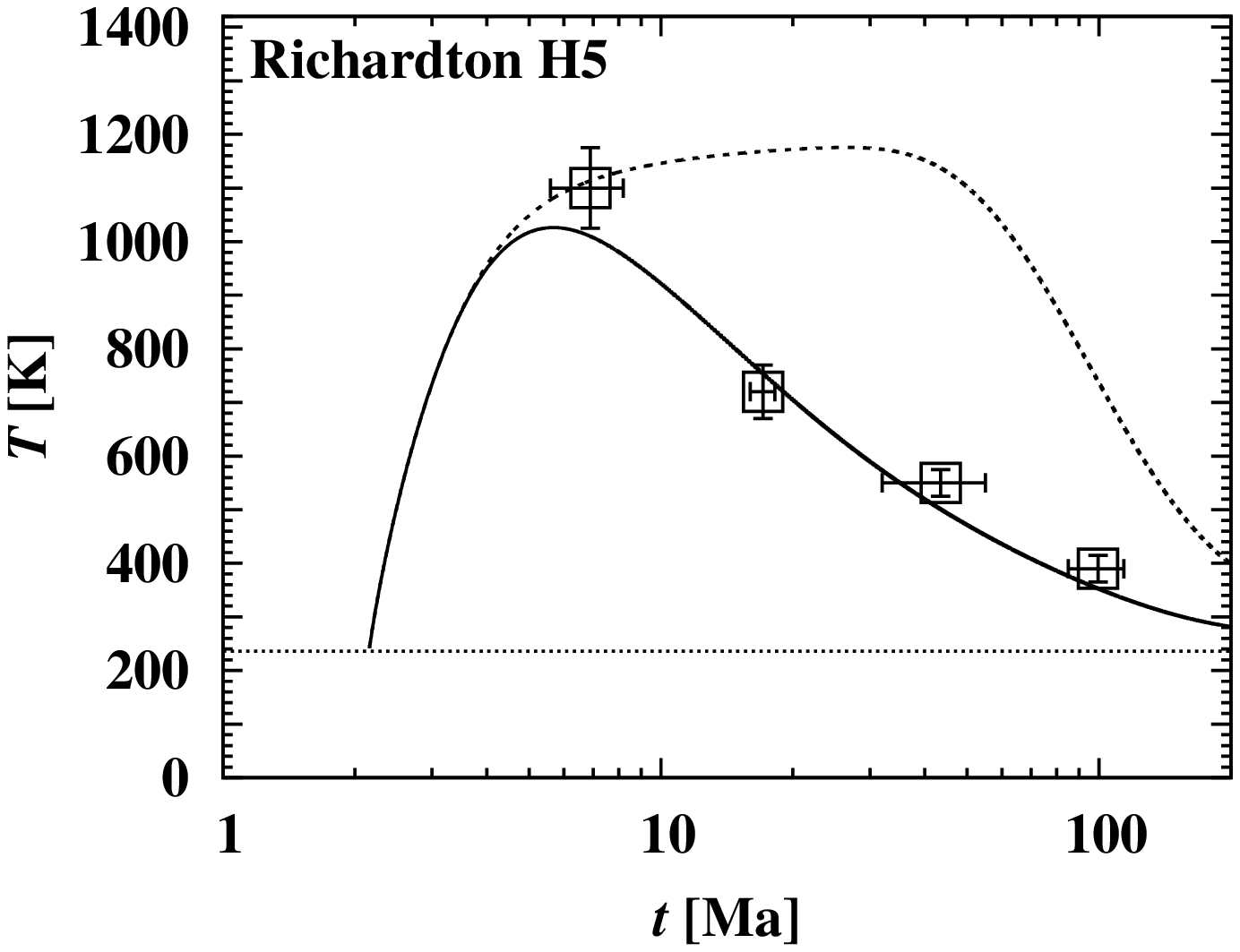}
\hfill
\includegraphics[width=0.31\hsize]{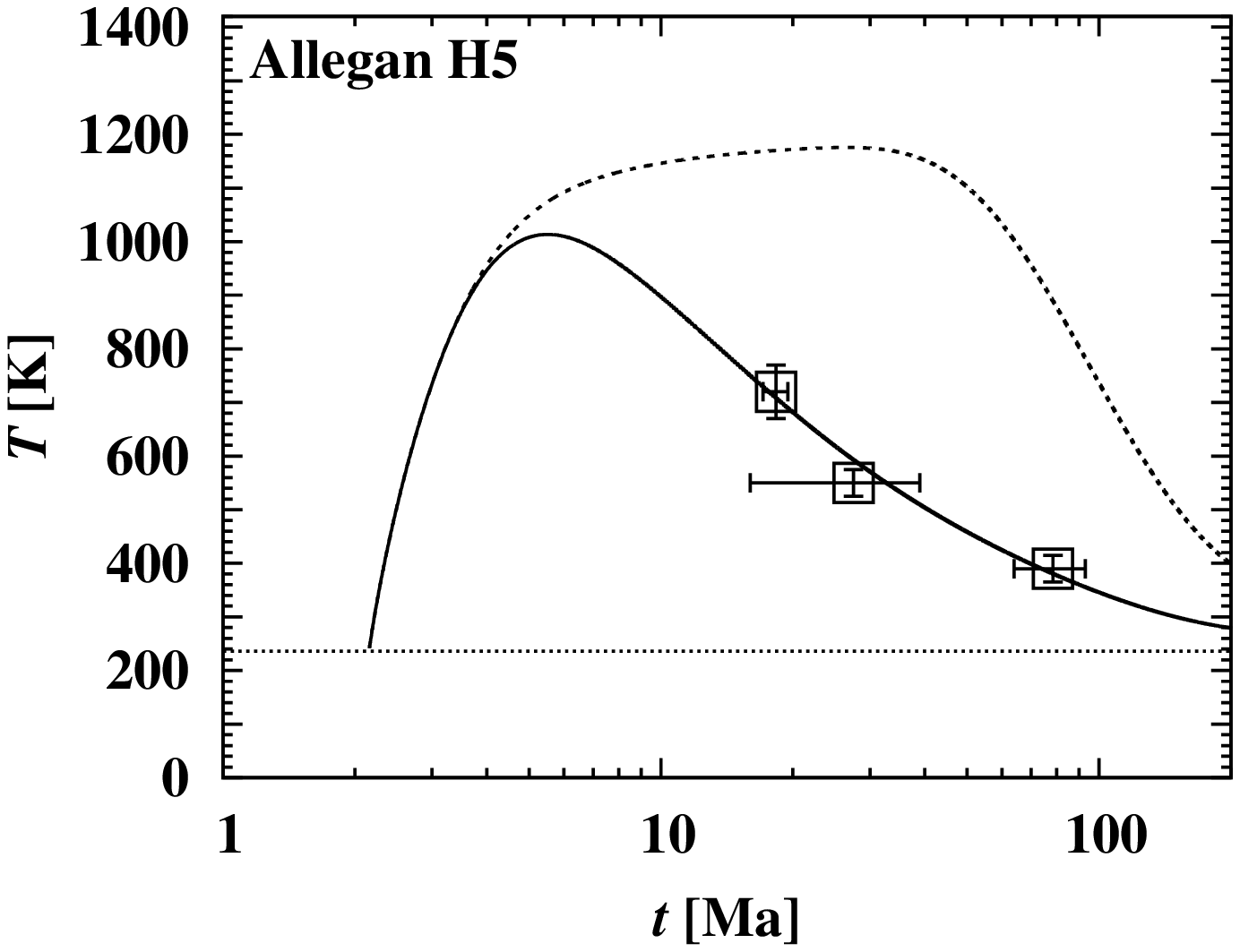} 
\hfill
\includegraphics[width=0.31\hsize]{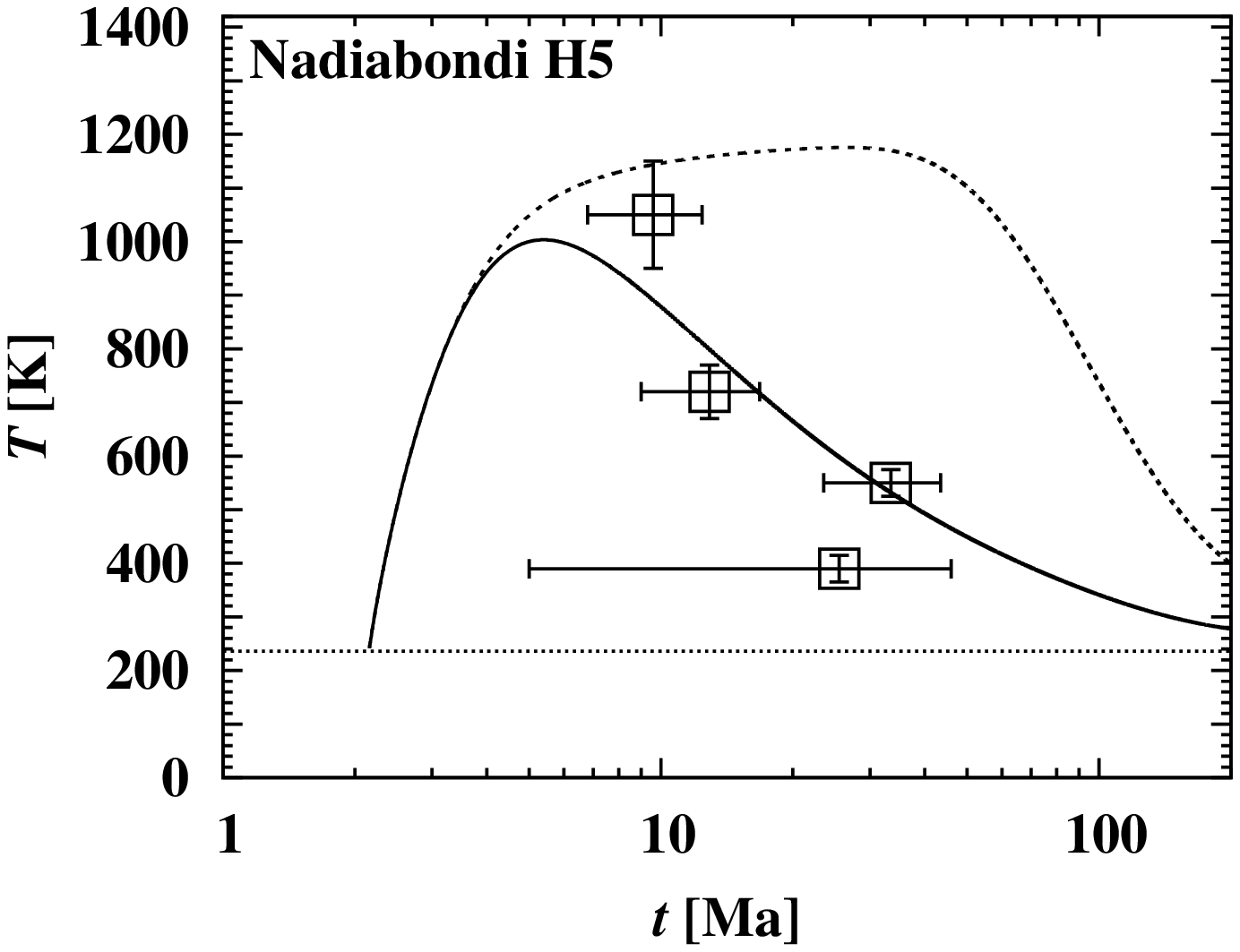} 
}

\centerline{
\includegraphics[width=0.31\hsize]{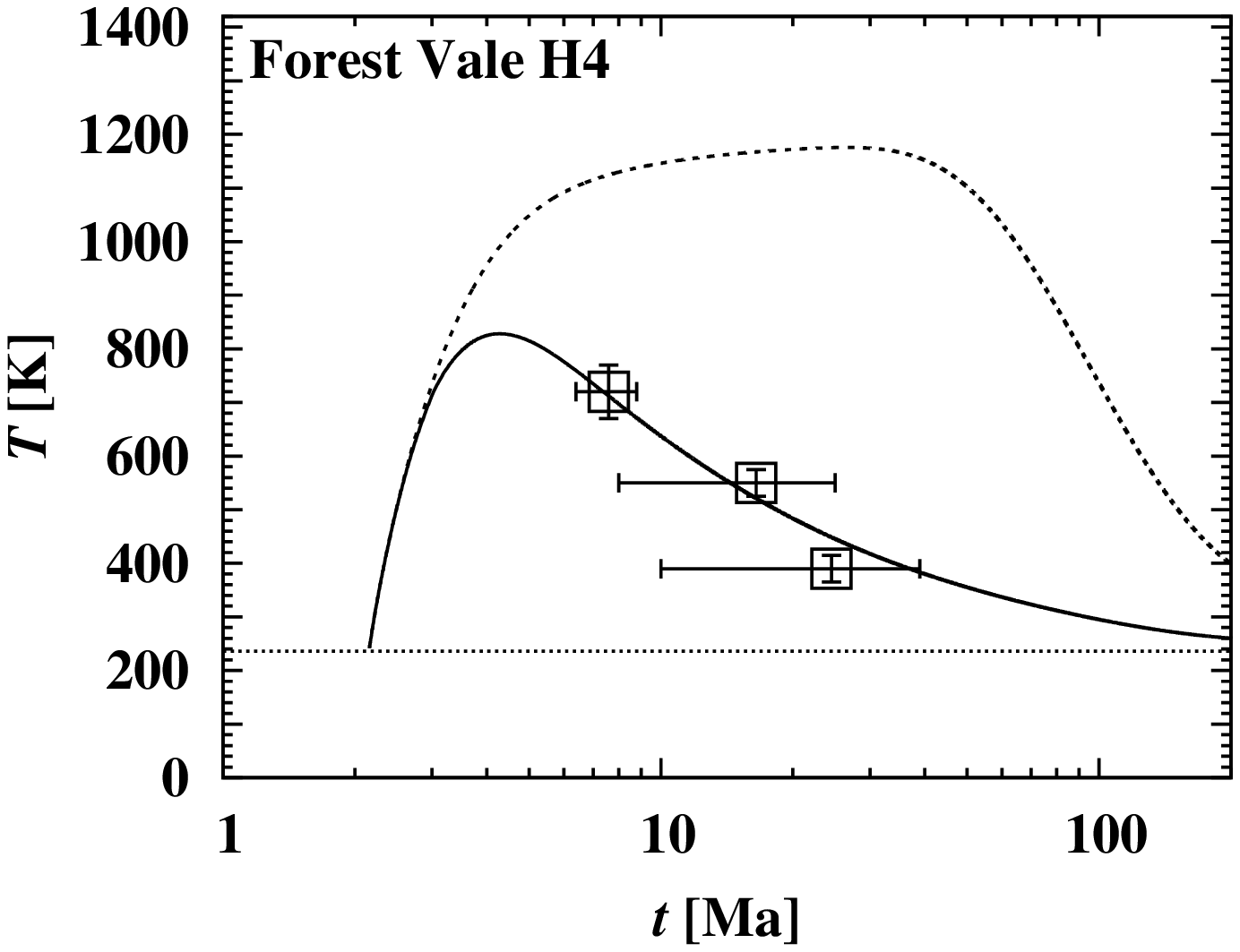}
\includegraphics[width=0.31\hsize]{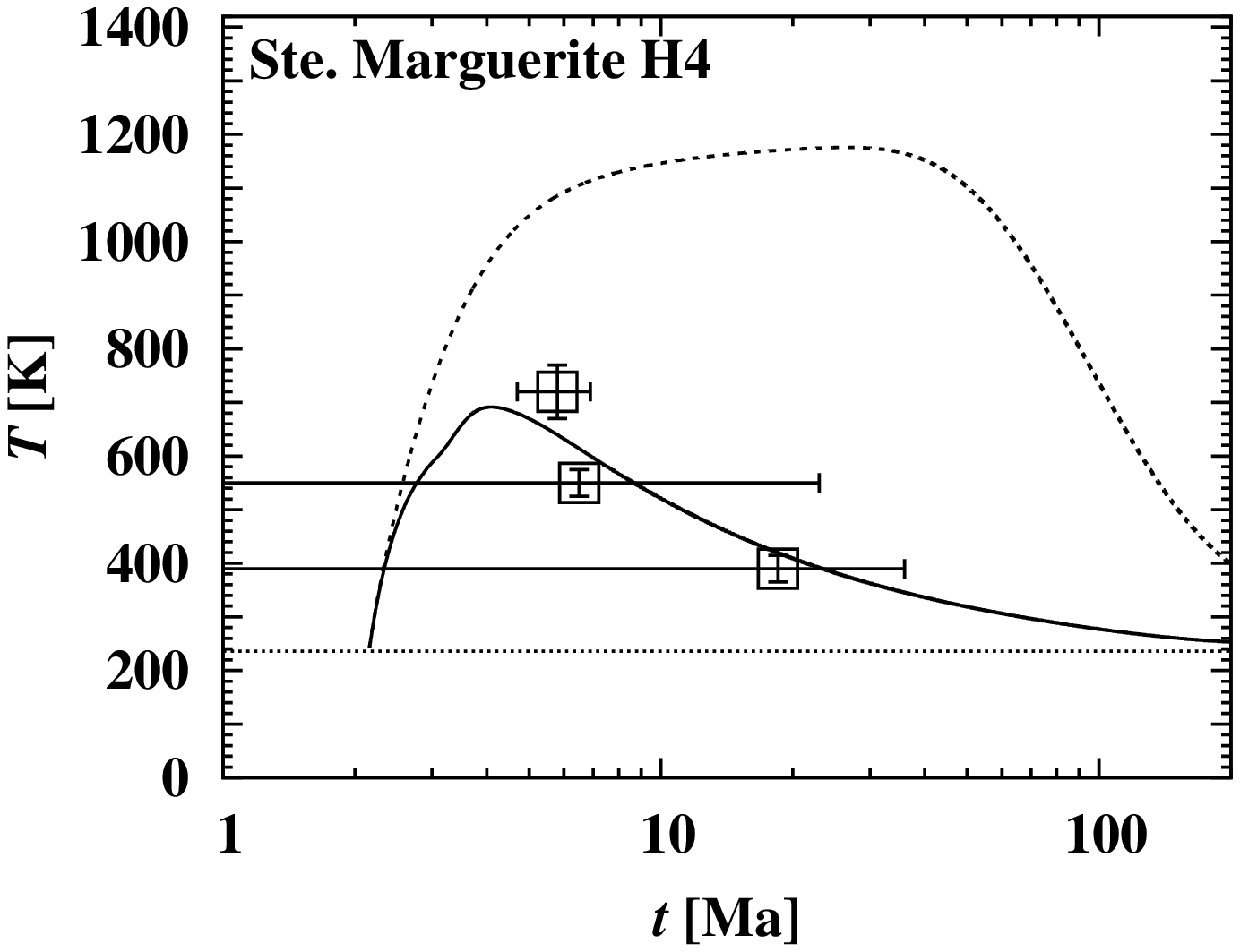}
}

\caption{Results for the individual meteorites if a Miyamoto-like model is 
used for the model fit. The burial depth's of the meteorites found for this
model are given in Table~\ref{TabDeptsMiya}. For meaning of symbols see caption
to Fig.~\ref{S_Evo1}.}

\label{S_Evo1Miya}
\end{figure*}

The thickness of the remaining porous outer layer is also given in
Table~\ref{TabDeptsVarK}. This layer is defined as that region where 
$\phi(r)\ge0.1$; its lower boarder is given by the radius where $\phi(r)=0.1$.
Around this border the heat conductivity $K(\phi)$ drops nearly discontinuously from
the bulk value $K_{\rm b}$ to the low value $K(\phi_{\rm srf})$. The blanketing
effect of this layer is responsible for the shallow burial depths of the H4 and
H5 chondrites found in the model, which are also found in other models 
\citep[e.g][]{Ben96,Akr98,Har10} that introduce an insulating outer regolith
layer instead of modelling the sintering process.

\begin{table}

\caption{%
Parameters and burial depths of the meteorites for the Miyamoto-like optimised 
H chondrite parent body model.}

\begin{tabular}{llllll}
\hline\hline
\noalign{\smallskip}
Quantity      & value &  unit \\
\noalign{\smallskip}
\hline 
\noalign{\smallskip}
\multicolumn{1}{l}{Radius}            & $R$ & 143.67 & km \\
\multicolumn{1}{l}{Formation time}    & $t_{\rm form}$ & 2.157 & Ma \\
\multicolumn{1}{l}{Heat conductivity} & $K_{\rm b}$ & 4.00 & W\,m$^{-1}$K$^{-1}$   \\
\multicolumn{1}{l}{$^{60}$Fe/\,$^{56}$Fe ratio} & & $2.07\times10^ {-7}$ & \\
\multicolumn{1}{l}{Surface temperature} & $T_{\rm srf}$ & 236.2 & K \\
\multicolumn{1}{l}{Porosity} & $\phi_{\rm srf}$ & 8.0\% & \\
\noalign{\smallskip}
\multicolumn{1}{l}{Maximum temperature} & $T_{\rm c}$ & 1\,175 & K \\
\multicolumn{1}{l}{Porous outer layer}  &              & none & km \\
\noalign{\smallskip}
\hline 
\noalign{\smallskip}
Meteorite      & type & \multicolumn{1}{c}{depth [km]} & \multicolumn{1}{c}{$T_{\rm max}$ [K]} \\
\noalign{\smallskip}
\hline 
\noalign{\smallskip}
Estacado       & H6   &   65.7             & 1\,150\\
Guare\~na      & H6   &   63.5             & 1\,149\\
Kernouv\'e     & H6   &   51.6             & 1\,136\\
Richardton     & H5   &   21.5             & 1\,026\\
Allegan        & H5   &   20.1             & 1\,013\\
Nadiabondi     & H5   &   18.8             & 1\,003\\
Forest Vale    & H4   &   \phantom{3}8.29  & \phantom{1\,}828\\
Ste. Marguerite& H4   &   \phantom{3}3.98  & \phantom{1\,}691\\
\noalign{\smallskip}
\hline
\end{tabular}

\label{TabDeptsMiya}
\end{table}

In this model the radius of the body is smaller than in the model with 
prescribed $K_{\rm b}$. With $R=123$\,km it is 30\,km bigger than the
average radius of 93\,km of (6) Hebe. If (6) Hebe really is the parent body, this would mean that material corresponding {\rm bf to 24\% of the initial radius or 
57\% of the initial mass has been knocked-off by collisions. This model seems to
be more plausible because the body then retained nearly half of its initial 
mass} and did not lose most of its mass as in the preceding model. This makes
it more plausible that we still find H4 chondrites. 

The value for $K_{\rm b}$ required for an optimum fit is significantly lower
than the average value for $K_{\rm b}$ obtained from the laboratory data but 
still inside the range of experimental values. This is shown in 
Fig.~\ref{FigVarKb} where we plot the ratio $K_{\rm exp}/K(\phi)$ for all 
chondrites from \citet{Yom83}. If one would have used only the experimental
values for the H chondrites, the optimised value for $K_{\rm b}$ would be 
definitely too low. This may result from the fact that, as already mentioned,
the H~chondrites used by \citet{Yom83} for measuring the heat conductivity and
its dependence on $\phi$ are unfortunately not representative for the parent
body as a whole. For instance could the material in the more central part of the
body have for some reason different properties than the zones probed by the 
meteorites. Another probability could be that we neglected the temperature
variation of the heat conductivity. If one looks at Fig.~\ref{FigVarKb} this
variation seems to be small, but in fact the heat conductivity drops with
increasing temperature and may be significantly lower at 1\,000\,K than the 
values seen in Fig.~\ref{FigVarKb}. Therefore the low value found in the 
optimisation may also simply mean that if we optimise $K_{\rm b}$, the
resulting value is kind of a ``best average'' between the lower conductivity of 
the hot core region of the parent body and of the higher conductivity in the 
cooler mantle region. The temperature dependence of the heat conductivity
therefore should be considered
in future model calculations.

\subsection{%
Model with given Radius of (6) Hebe}

We also tried a model fit with the present-day radius $R=93$\,km of (6)~Hebe
to check whether a somewhat sub-optimal but, nevertheless, acceptable fit could
be obtained for this case. The value of $K_{\rm b}$ is included in the 
optimisation process, but $R$ is now fixed. The model, again, used 250 
mass-shells. The optimisation was run for about 5\,500 generations with hundred
individuals per generation. The finally accepted optimum fit had a value of 
$\chi^2=6.543$ in generation 1\,633. With this value for $\chi^ 2$ the fit is
not worse than the fit  of the model with fixed $R$ in view of the fact that an
acceptable fit would require only $\chi^2\lesssim15$ (cf. 
Sect.~\ref{SectApplOpt}). The resulting set of parameters is given in 
Table~\ref{TabDeptsHebe} and the temperature variation at the burial depths of
the meteorites is shown in Fig.~\ref{FigOptHebe}.

The main difference between the models with optimised $R$ and fixed $R$ is that
the resulting $T_{\rm srf}$ in the first model is 225.0 and 306.5 K in the
latter, all other things being very similar. The higher surface temperature is
required to prevent a too rapid cooling of the outer layers of the smaller body.
In practice this means that there is some ambiguity in the models with respect
to the location of the body (surface temperature should decrease with increasing
distance to the sun) and its radius. The optimisation process cannot really
discriminate between the two cases considered here. 

\subsection{%
Nadiabondi and Ste. Marguerite}

For Nadiabondi and Ste. Marguerite, some data points need to be discussed
separately. 

For Ste. Marguerite the temperature increase during metamorphism was 
insufficient for resetting some of the radioisotopic systems, making it
difficult to constrain the cooling history of this meteorite. For instance,
\citet{Kle08} argued that the closure temperature of the Hf-W system is much 
higher than the metamorphic peak temperature of Ste. Marguerite. The Hf-W age of
$1.7\pm0.7$\,Ma after CAI formation, therefore, must date a high-temperature
event prior to the metamorphism. This event most likely was chondrule formation.
This interpretation is consistent with our finding that the H chondrite parent
body probably accreted at about 1.88\,Ma after CAI formation. 

Second, some data of relatively poor precision do not really constrain the fit
procedure, e.g., in the case of Nadiabondi the low temperature datum of fission
track retention of the phosphate merrillite. Nominally it yields an - impossible
- negative cooling rate, and only within errors just relatively fast cooling can
be inferred. Similarly, both fission track and Ar-Ar data of Ste.~Marguerite
have large uncertainties, precluding firm constraints on the cooling history,
although the nominal values plot on the fitted cooling-curves.

\subsection{%
A Miyamoto-like model}

For comparison, the optimization was also done for a model where the porosity was set to a value of $\phi_\mathrm{srf}=0.08$. This is a typical value
as it is observed for various H6 chondrites that have not been subject to strong shock modifications \citep{Yom83,Con08}. With this initial porosity the body is
practically already compacted from the beginning on and sintering does not play
any role. The heat conductivity then is practically constant throughout the body
and for all times. The prefactor $K_{\rm b}$ of the heat conductivity was set to
the same value as in Sect.~\ref{SectModelKfix} and was not included in the 
optimisation process. Such a model corresponds essentially to the kind of model 
onsidered originally by \citet{Miy81}, the mayor difference being that heating
by $^{60}$Fe is included.

The model, again, used 250 mass-shells. The optimisation was run for about 
3\,260 generations with hundred individuals per generation. In generation 2299
generations an optimum value of $\chi^2=5.94$ was obtained that was considered
as sufficiently good. The model parameters of the optimised model are shown in
Table~\ref{TabDeptsMiya} and  Fig.~\ref{S_Evo1Miya} shows the fit for the individual meteorites. 

It is obviously possible to obtain also an acceptable fit to the observed cooling history of the meteorites with such kind of model. The parameters of 
such a model are somewhat different from the preceding cases because of the lack
of an insulating porous outer layer and a higher heat conductivity during the
initial warm-up phase. A significant contribution of $^{60}$Fe decay to the heating is required to compensate for the more easy heat-loss in 
the present case. Though the model as a whole does not look worser than the
model with optimised  $K_{\rm b}$, it is physically not acceptable since
the lack of a porous outer layer is unrealistic.

\begin{table*}
\caption{%
Properties of the parent body of H chondrites derived by different
thermal evolution models, and some key data used for the modelling. In some
models an insulating outer regolith layer is assumed; in that cases a core radius
and layer thickness are given.}

\begin{tabular}{llccrclccll}
\hline\hline
\noalign{\smallskip}  
Model & \multicolumn{1}{c}{$R$}  & $t_{\rm form}$ & $T_{\rm srf}$ &\multicolumn{1}{c}{$M$} & $K$ & \multicolumn{1}{c}{$c_p$} & $\rho$ & $h^{\rm f}$ &Method & Sintering\\
      & \multicolumn{1}{c}{km} & Ma & K & \multicolumn{1}{c}{kg} & W/mK & \multicolumn{1}{c}{J/kgK} & kg/m$^3$ & W/kg & & \\
\noalign{\smallskip}
\hline
\noalign{\smallskip}
\citet{Miy81}  & 85   & 2.5 & 200 & $8.2\,10^{18}$ & 1 & 625 & 3\,200 & 
  $1.82\,10^{-7}$ & analytic & --- \\
\citet{BeS96}  & 99.0+1.0 & 2.2 & 200 & n.a.$^{\rm e}$ &  variable$^{\rm d}$  & variable$^{\rm b}$ &        3\,780$^{\rm a}$ & n.a.$^{\rm e}$
& numeric & $\surd$ + regolith \\
\citet{Akr98} & 97.5+2.5 & 2.3 & 200 & $1.4\,10^{19}$ & variable$^{\rm d}$  &variable$^{\rm c}$ & 3\,450 &  &  numeric & regolith 
 \\
\citet{Tri03}  & 100 & 2.5 & 300 & $1.3\,10^{19}$ & 1 & 625 & 3\,200 & 
 $1.82\,10^{-7}$ & analytic & --- \\
 \citet{Kle08} & 100 & 2.7 & 300 & $8.2\,10^{18}$ & 1 & 625 & 3\,200 & 
  $1.82\,10^{-7}$ & analytic & --- \\
\citet{Har10} & 99.2+0.8 & 2.2 & 170 & $1.4\,10^{19}$ & variable$^{\rm d}$  & variable$^{\rm b}$ & 3\,250 
& $1.82\,10^{-7}$ & numeric &  regolith \\
This work & 123 & 1.88 & 225 & $3.0\,10^{19}$ & 2.65$^{\rm a}$ & 
variable$^{\rm b}$ & 3\,780$^{\rm a}$ & $1.67\,10^{-7}$ & numeric & $\surd$ \\
\noalign{\smallskip}
\hline
\noalign{\smallskip}
(6) Hebe$^{\rm g}$ (today) & 93 & & & $1.28\,10^{19}$   & & & 3810 & \\
   \multicolumn{2}{r}{(diameter $205\times185\times170$ km)}  & & & $\pm6.4\,10^{17}$ & & & $\pm260$ & \\
\noalign{\smallskip}
\hline
\end{tabular}

\medskip\noindent
{\scriptsize
Notes: (a) for consolidated material, varies with porosity 
(b) fit of $c_p$ to meteoritic data from \citet{Yom84}, at $T=500$ the value 
is 865\,J/kgK 
(c) forsterite,
(d)  fit of $K$ to meteoritic data from \citet{Yom84}, at $T=500$ the value of the
consolidated material is  3.66\,W/mK and varies only slightly with temperature, 
(e) no or insufficient data presented,
(f) at time of CAI formation, given value for $^{26\!}$Al only, some models
consider also the less efficient heat sources ($^{60}$Fe, long lived radioactives),
(g) from data collection \citet{Bae11}.
}

\label{TabBodyParm}
\end{table*}

\subsection{%
Comparison with earlier results}

There are a number of previous attempts to reconstruct the properties of the
H chondrite parent body using the empirically determined cooling rates of H chondrites derived from
different depth within the parent body.
Table \ref{TabBodyParm} gives an overview on the results and the values of some
key parameters of some of these models. The table
lists all the models that attempted to obtain in some way
a ``best fit'' between properties and cooling histories derived from laboratory
studies and a thermal evolution model of the parent body. The results are
generally the putative radius and formation time of the body, and the burial
depths of the meteorites. All studies, including ours, showed, that
the observed properties of the H chondrites are compatible with a parent body 
of about 100\,km radius and an ``onion shell'' model. In this model the body is
first heated within a period of ca. 3\,Ma by decay of radioactive nuclei, mainly
$^{26\!}$Al, and then cools down over an extended period of ca. 100\,Ma, without
suffering catastrophic collisions that would have disrupted the body, thereby disturbing its thermal evolution record.

Despite the different assumptions made in the model calculations, from the
rather simplistic model of a homogeneous body with constant material properties,
that can be treated analytically \citep{Miy81}, to models with rather complex
implemented physics like that of \citet{Har10} and ours, the derived properties
of the parent body are similar. These do not seem to depend very critical on the
particular assumptions. The more strongly deviating results for the burial
depths of the meteorites cannot be used to discriminate between models, because
no independent information on this is presently available. This would require
that methods are developed to determine the maximum pressure experienced by a
meteorite. 

The most important difference between our model and previous models is that
we included sintering of the granular material in the modelling. This process is 
not considered in other models, except for \citet{Yom84} and  \citet{BeS96}.
The model of \citet{Yom84} unfortunately does not consider the important
contribution of $^{26\!}$Al to the heating. Therefore this model cannot be
compared to other models. \citet{BeS96} gave only very few details on their
model results such that one cannot make a comparison with our and other models
also in this case. The models of \citet{Akr98} and \citet{Har10} do not consider
sintering, but simulate the insulating effect of a porous outer layer, that
escaped sintering, over a consolidated body by introducing two outer layers
with low filling factor, a ``regolith'' and a ``mega-regolith'' layer. The 
thickness's of both layers were assumed by \citet{Akr98}, but they were
determined by an optimisation of the model by \citet{Har10} and  a thickness
of 410\,m was obtained for each of the layers. This compares well with a 
thickness of $<1$\,km  of a residual porous outer layer in our model. The
thermal shielding of the consolidated interior of the body by a thin insulating
outer layer of low thermal conductivity is obviously essential for obtaining a
good fit between thermal evolution models to meteoritic data as it is obtained
in the model by \citet{Har10} and in ours. 

A major difference between our model results and others is the earlier derived
formation time, $t_{\rm form}$. This difference probably results from the lower 
mass-density used in other model calculations and the resulting lower mass of
the body. Because of the resulting higher total heat capacity of the whole body 
($\int\!\rho c_{\rm p}{\rm d}V$) in our model more heat from decay of 
radioactive nuclei is required to heat the body to the temperatures as witnessed
by the observed cooling history of the various meteorites. Hence, an earlier
formation time is required for our optimised model since the specific heating
rate at zero time (equal to time of CAI formation) is almost the same in all models. The
higher average density in our model is a result of the sintering process which
results in very low or vanishing porosity over most of the volume of the body.

If (6) Hebe really is the parent body,  the mass-density in our model fits well
to that determined for Hebe \citep{Bae08,Bae11}. The densities used in other
model calculations then would be too low and the earlier formation time found
in the present model calculation therefore is probably more realistic than the 
later formation times found in other models. 

The surface temperature, $T_{\rm srf}$, is determined in our model by the
optimisation process, in the other model it seems to be chosen more or less 
arbitrary within the limit expected for bodies in the asteroid belt and in some
cases adjusted to improve the model fit between closing temperatures and closing
times observed for meteorites. The dependence of the thermal evolution model
on the value of $T_{\rm srf}$ is generally weak \citep[e.g.][]{Hen11}. The value 
determined by our optimisation procedure hinges mainly upon the fit to the two 
H4~chondrites that are most strongly affected by $T_{\rm srf}$ because of the
shallow burial depths required by their petrologic type. 

There are a number of other model calculations \citep{Ben96,Hev06,Sah07,War11}
that also studied the thermal evolution of asteroids and give important insight
into their properties and the processes operating in them, but did not attempt 
to obtain definite results for the properties of the parent body of H chondrites
from some kind of ``best fit'' to meteoritic data. Only the range of possible
values is studied in these papers. This does not enable a direct comparison, but
the range of possible values for radii and formation times given are in accord
with the results shown in Table~\ref{TabBodyParm}.

\subsection{Orbit of parent body}

If the number obtained for $T_{\rm srf}$ from our optimisation process is taken
seriously, it measures the average surface temperature for the first 
$\approx20\dots30$\,Ma of evolution where the temperature for Forest Vale and
Ste.~Marguerite was above the closure temperature for Pu fission tracks. Since 
the observed lifetime of accretion disks is less than 10\,Ma \citep{Hai01} 
the surface temperature is determined by illumination by the young proto-sun and
thermal emission from the surface, in particular during the most critical phase
where the temperature at the burial depth evolved through the closing temperature
for Pu fission tracks. Then we can estimate the average radius of the orbit of
the parent body from the relation (for a rapidly rotating body)
\begin{equation}
\sigma T_{\rm srf}^4={(1-A)S(1)\over 4a^2}\,,
\end{equation}
where $\sigma$ is the Stefan-Boltzmann constant, $A$ the albedo of the surface,
$a$ the distance to the protosun in AU, and $S(1)$ the solar constant at 1\,AU
at this early evolution period. 

For the solar constant $S(1)$ we have to observe that the sun ignites at
about 10\,Ma hydrogen burning and has the luminosity of a 1\,M$_{\sun}$
star on the {\rm zero age} main sequence. Its luminosity during the first 
100\,Ma is by 
about 30\% less than its present value \citep[e.g.][]{Dan94}. The typical 
surface albedo for S asteroids is 0.2 \citep{Rya10}. The surface albedo of 
(6) Hebe is particular high with $0.291\pm0.029$ \citep{Rya10}. The albedo found
today is probably not typical for the material that covered the surfaces of the
bodies 4.5\,Ga ago, because highly processed material and the products of its 
grinding are presented at the surface. It must had contained a lot of more 
fine-grained material that only later matured by crystal growth at elevated
temperatures and probably was darker. 

We assume arbitrarily $A=0.1$ and find a radius of the orbit of $a=1.23$\,AU (an 
albedo of $A=0.2$ yields  $a=1.16$\,AU). This is smaller than the present 
average radius of the orbit of $2.4$\,AU. If the temperature $T_{\rm srf}$ found
in optimisation is really representative for the average surface temperature
during the essential phase of the thermal evolution, and if (6) Hebe is the
parent body of the H chondrites, then our result means that (6) Hebe formed
closer to sun than its present distance. However, a more detailed treatment of
the processes determining the surface temperature is necessary, before any
definite conclusions can be drawn.

\section{Concluding remarks}
\label{SectConclu}

We combined the code for modelling the internal structure and evolution of 
planetesimals of the 100\,km size class \citep{Hen11} with a code for
optimisation of a quality function that is based on an evolution algorithm.
This allows finding a consistent set of parameters that describe the properties
and evolution of a planetesimal and the burial depths of the meteorites. A such
the thermal evolution of the model body reproduces very closely the empirically 
determined cooling history of all H chondrites with sufficient available data. 

The results show, that the ``onion shell'' model successfully reproduces the
thermal evolution history of all the meteorites used in this study. For six of 
the eight meteorites a burial depth can be found for which  the empirical cooling 
data can be fitted in an excellent way with the local evolution of temperature at
that burial depth. For the remaining two cases it is still possible to find a
burial depth that results in a good fit if the Pu fission-track datum of
Nadiabondi is rejected. This confirms earlier conclusions that the thermal evolution of the
 H chondrite parent body and its initial thermal structure has not been disturbed by catastrophic impacts. 

The reconstructed properties of the parent body of the H chondrites are
comparable to what has been found in other model calculations. The different
published models treated the problem in different kinds of approximations,
solving the heat conduction equation being almost the only feature common to
all models, and used different values for the material properties. Our
estimated size of 123\,km of the parent body is somewhat bigger compared to
the 100\,km found in other models, and requires a slightly earlier formation
time of 1.88\,Ma compared to the 2.2 -- 2.5\,Ma found in other models. Such
minor differences are certainly due to the different approaches used.

Our study shows, that the input physics of the models has to be improved,
before more reliable data on the parent body can be obtained. Such
improvements are currently underway in our group. In particular the heat
conductivity has to be calculated from the specific properties of chondritic
material. It will also be necessary to obtain more precise radiometric ages 
for a wider range of samples.

The advantage of our method is that it allows for a simultaneous and
consistent fit of several model parameters to a big number of observational
restrictions and is very flexible if the kind of data to be included in the
optimization process is to be changed. A minor shortcoming of the
optimisation method used is that a considerable number of complete evolution
models has to be calculated during the course of the optimisation process,
but our implicit solution method is fast enough that no restriction as to the
applicability of the method to our problem results from this.


\begin{acknowledgements}
We thank the referee S. Sahijpal for a constructive and helpful referee report. This work was supported in part by `Forschergruppe 759' and in part by
`Schwerpunktprogramm 1385'. Both are supported by the `Deutsche 
Forschungs\-gemeinschaft (DFG)'. S.H. is Member of the IMPRS for 
Astronomy \& Cosmic Physics at the University of Heidelberg.
\end{acknowledgements}


\end{document}